# Compaction of bacterial genomic DNA: Clarifying the concepts

Marc JOYEUX

*Laboratoire Interdisciplinaire de Physique (CNRS UMR5588),*

*Université Joseph Fourier Grenoble 1, BP 87, 38402 St Martin d'Hères, France*

Email : marc.joyeux@ujf-grenoble.fr

Abstract : The unconstrained genomic DNA of bacteria forms a coil, which volume exceeds 1000 times the volume of the cell. Since prokaryotes lack a membrane-bound nucleus, in sharp contrast with eukaryotes, the DNA may consequently be expected to occupy the whole available volume when constrained to fit in the cell. Still, it has been known for more than half a century that the DNA is localized in a well defined region of the cell, called the nucleoid, which occupies only 15% to 25% of the total volume. Although this problem has focused the attention of many scientists for the past decades, there is still no certainty concerning the mechanism that enables such a dramatic compaction. The goal of this Topical Review is to take stock of our knowledge on this question by listing all possible compaction mechanisms with the proclaimed desire to clarify the physical principles they are based upon and discuss them in the light of experimental results and the results of simulations based on coarse-grained models. In particular, the fundamental differences between $\psi$-condensation and segregative phase separation and between the condensation by small and long polycations are highlighted. This review suggests that the importance of certain mechanisms, like supercoiling and the architectural properties of DNA-bridging and DNA-bending nucleoid proteins, may have been overestimated, whereas other mechanisms, like segregative phase separation and the self-association of nucleoid proteins, as well as the possible role of the synergy of two or more mechanisms, may conversely deserve more attention.

# 1 – Introduction

Illustrations of the hierarchical compaction of genomic DNA in the nuclei of eukaryotic cells can be found in any textbook, from the initial wrapping of DNA around histone proteins to the final X-shaped chromosomes, through the various levels of fiber condensation. While seemingly simpler, the compaction of bacterial genomic DNA is nevertheless more poorly understood. One of the main reasons is that typical cell dimensions and DNA size of prokaryotes are significantly smaller than those of eukaryotes, with cell radii of the order of 1 µm against 10 to 100 µm and DNA size of the order of millions of base pairs against billions of base pairs. As a consequence, optical microscopy experiments are able to show that DNA occupies only a small part of the cell (from 15% to 25%) but fail to provide more detail because of resolution issues [1]. In contrast, electronic microscopy is able to provide information on the ultra-structure of the nucleoid (the region where the DNA is localized) but results depend dramatically on the experimental procedure that is used to prepare the cells [1,2]. Finally, the more recent techniques that consist in labeling specific genes with fluorescent dyes or proteins [3] usually provide information on the dynamics close to the loci of these genes but not on the global organization of the nucleoid. Owing to these difficulties, even seemingly simple questions lack an answer, despite the long lasting efforts of many groups, including those of Cozzarelli [4], Higgins [5], Murphy and Zimmerman [6], Kornberg [7], Woldringh [8], Dame [9], Busby [10], Shapiro [11], Austin [12], Boccard [13], Sherratt [3], and Kleckner [14] (this is but a very partial list of the groups involved in the characterization of the bacterial nucleoid, and an even more partial selection of their results, aimed at illustrating the diversity of the investigated problems and of the techniques used to tackle them). Among the unsolved problems, the compaction of DNA inside the nucleoid is particularly challenging. Estimations based on models like the Worm-Like-Chain indicate that, because of the rigidity arising from the bending energy term, unconstrained bacterial DNA molecules form a coil with a volume at least 1000 times that of the cell. Since the nucleoid is not delimited by a membrane, in sharp contrast with the nucleus of eukaryotic cells, it may naively be expected that the DNA would occupy the whole available volume when constrained to fit in the cell. Then, why does the nucleoid occupy only about 15% [15] to 25% [16] of the cell, as is for example clearly seen in figure 1 ? The mechanism that enables such a dramatic compaction has intrigued scientists for the past decades. It has been argued [17] that four essential mechanisms may contribute to the compaction of bacterial DNA inside the nucleoid, namely (i) the association of Nucleoid Associated Proteins with DNA, (ii) DNA supercoiling resulting from the over- or under-winding of a DNA strand, (iii) macromolecular crowding owing to the cytoplasm, and (iv) neutralization of the charges carried by the DNA molecule by multivalent ions and certain DNA binding proteins, but the importance of each mechanism and their possible synergies are still the matter of on-going debate. Moreover, it appears that there is sometimes some confusion in the mechanisms that are evoked to explain new experimental observations. For example, $\psi$-condensation (see below) is very often mentioned as the probable cause for the observed compaction, although this mechanism displays some peculiar properties, which should in certain cases rule it out of the list of possible explanations. This Topical Review focuses precisely on listing all possible compaction mechanisms, with the proclaimed desire to clarify the physical principles they are based upon, to discuss them in the light of recent experimental results, and to illustrate them as far as possible with results of simulations based on coarse-grained models. These models, where up to 15 DNA base pairs may be represented by a single site and proteins by a few ones, lack most of the details of atomistic models and are not appropriate for investigating specific interactions, but they are sufficient when non-specific interactions (like electrostatic ones) play the key role and they allow for the numerical integration of quite long trajectories for rather large systems. They will be used in this Review to associate explicit numbers with each mechanism and provide clear illustrations thereof.

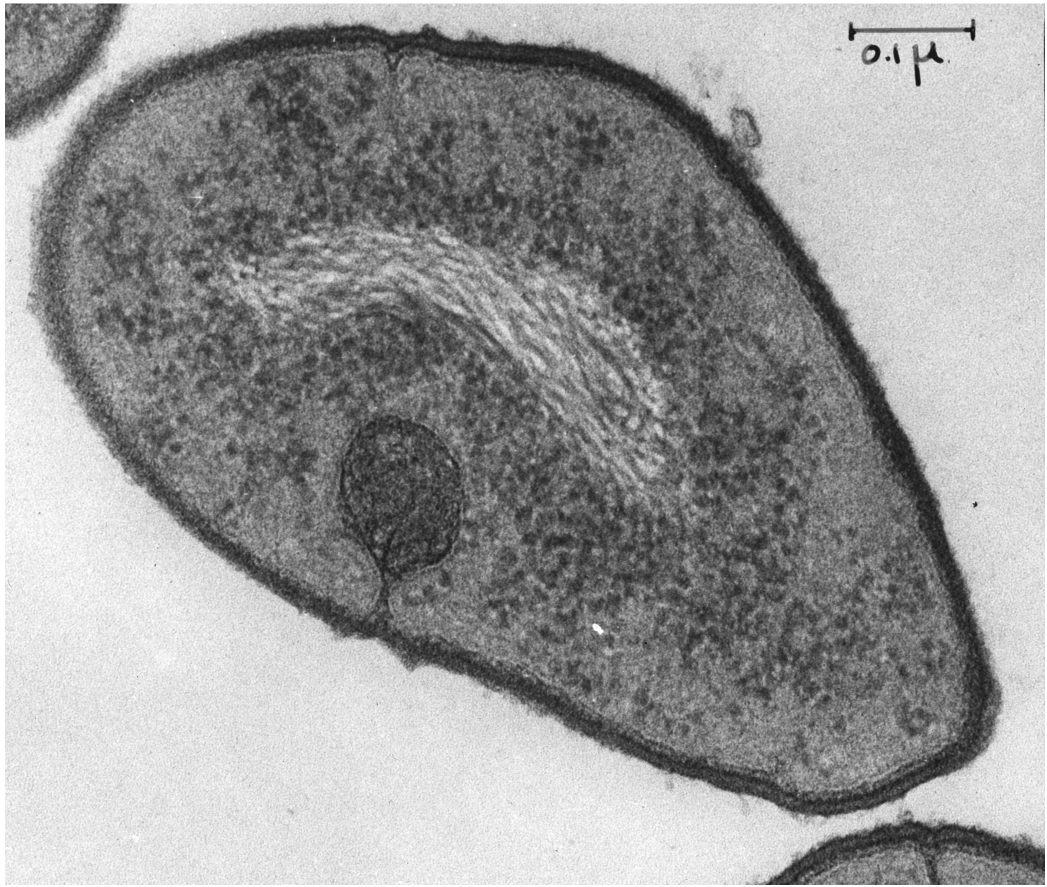

**Figure 1** : Transmission electron micrograph of a thin section through the bacterium *Diplococcus pneumoniae*. The fibrous nucleoid is centrally located, surrounded by cytoplasm containing numerous ribosomes. Micrograph recorded at a magnification of 64000x. Original 3.25 in. x 4 in. Photograph by George E. Palade and James Jamieson.
Link : http://www.cellimagelibrary.org/images/41048
Creative Commons Attribution, Non-Commercial Share Alike License :
http://creativecommons.org/licenses/by-nc-sa/3.0/legalcode

Mechanisms that are not specific to DNA, but may instead contribute to the compaction of any charged polymer (associative and segregative phase separation, correlation forces, $\psi$-condensation), are described in Sect. 2, while those which are more specific to DNA (supercoiling, bridging by Nucleoid Associated Proteins) are discussed in Sect. 3. Full detail of the models is provided in the Appendix.

**2 – Compaction mechanisms generic to all charged polymers**

DNA is a highly charged polymer. Owing to the $PO_4^-$ groups that alternate with deoxyribose ones along each strand, its linear charge is $-2e$ per base pair (where $e$ denotes the absolute charge of the electron), that is approximately $-e$ per 0.17 nm since two successive base pairs are separated by about 0.34 nm. Therefore, DNA usually comes as a salt, with $K^+$ or $Na^+$ cations as counterions, the charge neutrality for a 4.6 Mbp genomic DNA confined in a 1.5 $\mu m^3$ cell (like for *E. coli*) being achieved for a concentration of monovalent cations close to 0.01 M (the total physiological salt concentration is close to 0.1 M). The cytoplasm also contains large amounts of divalent metal ions, like $Mg^{2+}$ or $Ca^{2+}$, which are essential for the activity of many enzymes [18,19]. The cytoplasm consequently behaves like a polyelectrolyte, a state of condensed matter that is much more poorly understood than neutral polymer solutions. Difficulties arise principally because of the long range nature of Coulomb interactions. For example, computer simulations [20,21] and theory [22] have pointed out that for most dilute and semi-dilute solutions containing long charged polymers part of the counterions are trapped in the volume of the macromolecular coil, while the other part escapes to the remainder of the accessible volume. The DNA molecule therefore retains a nonzero charge even in salt solutions, so that electrostatic repulsion tends to oppose DNA compaction. This is the first point that should be kept in mind, namely that the *principal* *force that opposes the compaction of genomic DNA in the nucleoid is the electrostatic repulsion between the charged duplexes*. Assuming a reduction to about 10% of its original value of the effective charge on the DNA due to charge neutralization by the counterions (see below), the free energy change due to electrostatic repulsion upon compaction of the DNA in the nucleoid has been estimated to be still as large as about 0.2 $k_B T$ per base pair [23].

The second point one should keep in mind is that *a large number of proteins and other macromolecules are present in the cell, in addition to the DNA and the small counterions*, including typically around 200-300 mg of proteins and 100 mg of RNA per ml of cytoplasm in *E. coli* cells [24]. Some of these molecules, like the RNA and the polyamine spermidine$^{3+}$, which is present in the mM range in bacteria, are multivalent ions that contribute to both macromolecular crowding and electrostatic balance.

In this section, I will review possible DNA compaction mechanisms, which rely uniquely on the charged polymer character of the DNA. These mechanisms may be separated into two groups, according to the strategy they use to overcome the DNA-DNA electrostatic repulsion. Mechanisms of the first group are based on strong electrostatic interactions between DNA and the other macromolecules in the cytoplasm, while mechanisms of the second group require instead that the charges on the DNA be neutralized as completely as possible to allow additional weak forces to induce DNA compaction. This distinction is admittedly somewhat superficial, because the charges located on the phosphate groups of the DNA are always partly neutralized by the cloud of companion cations, but it will be shown below that it still helps discriminating between rather different physical mechanisms.

**A – Phase separation induced by strong electrostatic interactions.**

As a matter of fact, solutions containing different types of macromolecules may spontaneously undergo phase separation above certain concentrations, leading to phases with different concentrations of the solutes and different physico-chemical properties [25]. Since the cytoplasm contains high concentrations of nucleic acids and proteins, it has been suggested that the formation of the nucleoid may be an illustration of such phase separation, with one phase (the nucleoid) containing large amounts of DNA and the other phase (the cytoplasm) containing only small amounts thereof [26]. In Ref. [26], this reasoning was pursued one step further and it was tentatively proposed that the micro-compartmentation observed in the cytoplasm [27,28] may actually reflect the separation of the many macromolecular species found in the cytoplasm into more than two phases. Focusing on the case of two different types of macromolecules, say, A and B (with A-type molecules repelling other A-types molecules and B-type molecules repelling other B-type molecules), one may distinguish between two kinds of phase separation, depending on whether A-type and B-type molecules attract or repel each other. In the attractive case, which is usually observed when A-type and B-type molecules have segments that carry opposite net charges, the phase separation, which is described as *complex coacervation* or *associative phase separation*, leads to a dense phase that is rich in both A and B and a supernatant phase that is depleted in both A and B. Although less intuitive, the case where A-type and B-type molecules repel each other may also lead to phase separation, which is called *segregative phase separation*, because one of the resulting phases is rich in A and poor in B and *vice versa* for the other phase. The principal characteristics of phase separation in systems composed of two different polymers in solution are well explained by the Flory-Huggins theory of polymer solutions [29] and later developments thereof (see for example [30]) and the reader interested in the thermodynamical aspects of phase separation is prompted to refer to these works. In the description below, I will stay at a more qualitative molecular description of the occurring phenomena.

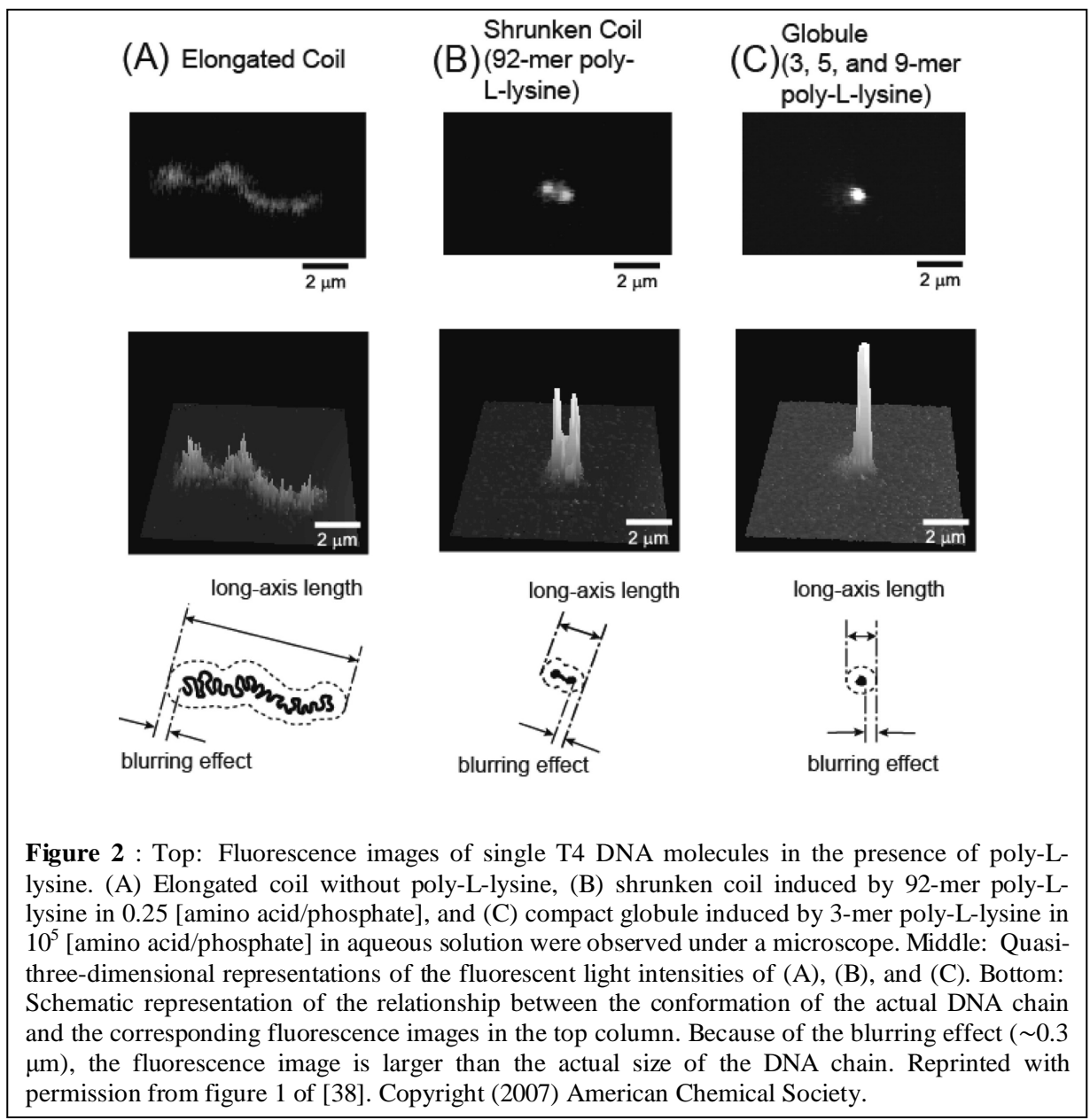

**Figure 2**: Top: Fluorescence images of single T4 DNA molecules in the presence of poly-L-lysine. (A) Elongated coil without poly-L-lysine, (B) shrunken coil induced by 92-mer poly-L-lysine in 0.25 [amino acid/phosphate], and (C) compact globule induced by 3-mer poly-L-lysine in $10^5$ [amino acid/phosphate] in aqueous solution were observed under a microscope. Middle: Quasi-three-dimensional representations of the fluorescent light intensities of (A), (B), and (C). Bottom: Schematic representation of the relationship between the conformation of the actual DNA chain and the corresponding fluorescence images in the top column. Because of the blurring effect (~0.3 μm), the fluorescence image is larger than the actual size of the DNA chain. Reprinted with permission from figure 1 of [38]. Copyright (2007) American Chemical Society.

*Associative phase separation (complex coacervation).*

As mentioned above, coacervates of DNA and other macromolecules may form when the latter ones have positively charged segments. This is precisely the case for cationic poly-amino acids, like poly-L-lysine. At neutral pH, each lysine residue has a positively charged $-NH_3^+$ amino group, so that sufficiently long polylysine molecules are able to bind several DNA duplexes simultaneously. Polylysines have been shown to be very efficient in condensing DNA into compact nanostructures [31,32,33] and are being used for *in vitro* and *in vivo* delivery of therapeutic DNA [34,35,36]. More precisely, it has been shown that 92-mer and 981-mer poly-L-lysine induce a gradual (*i.e.* continuous) compaction of DNA when increasing the ratio of the number of amino acids over the number of phosphate groups and that the compaction is maximum at a ratio of about 1 [37,38]. The shrunken coil, which is obtained for single T4 DNA molecules and 92-mer poly-L-lysine at a lysine/phosphate ratio of 0.25 is nicely illustrated in figure 2(B).

Coarse-grained modeling of DNA compaction by polycations of increasing length has been reported in [39]. In this work, DNA is modeled as a chain of 120 monovalent negatively charged beads separated at equilibrium by 0.5 nm and the polycations by several chains of varying length (3, 4, 5, 10, 15 and 30 beads) with monovalent positively charged beads separated at equilibrium by 0.56 nm. The number of positively charged beads is equal to that of negatively charged ones and the whole system is enclosed in a sphere of radius 30 nm. Successive beads along the same chain interact through stretching and bending energy terms and all beads interact via electrostatic terms including hard-sphere repulsion. The dynamics of the system was investigated through Monte Carlo simulations using the Metropolis algorithm. Typical conformations obtained for polycations with 3, 4, 5, 10 and 30 beads are shown in figure 3, where black beads represent the DNA and white ones the polycations. It is obvious from this figure that, for equal concentrations of positively and negatively charged beads, the ability to compact DNA increases markedly with the length of the polycations, with positive chains of 3 and 4 beads leading only to mild compaction and longer ones to quite compact complexes. There admittedly exist important differences between real DNA and the model proposed in [39]. For example, the persistence length of the model is only 1.2 nm, against about 50 nm for DNA. Moreover, the DNA chain is quite short and the concentrations of DNA and polycations are quite low. Still, this work has the virtue of emphasizing the crucial role of the length of the polycations for the purpose of compacting DNA. Experimental studies have confirmed the stronger compaction efficiency of longer polycations [37,40], but they also pointed out that the compaction mechanism of long polycation chains differs markedly from that of small ones [37,38]. We will come back to this point in section 2B.

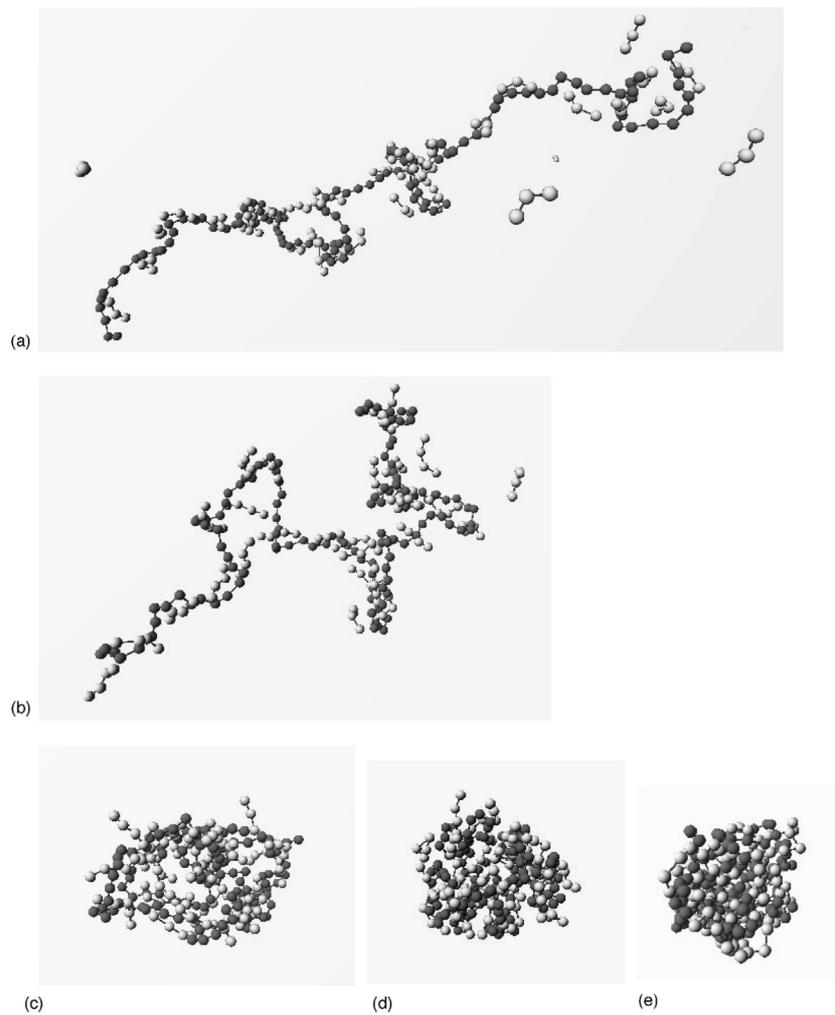

**Figure 3** : Typical configurations in systems with a polyelectrolyte with 120 negatively charged (black) beads and (a) 40 chains with 3 positively charged (white) beads, (b) 30 chains with 4 positively charged beads, (c) 24 chains with 5 positively charged beads, (d) 12 chains with 10 positively charged beads, (e) 4 chains with 30 positively charged beads. All beads are of unit charge. Reprinted with permission from figure 5 of [39]. Copyright 2003, American Institute of Physics.

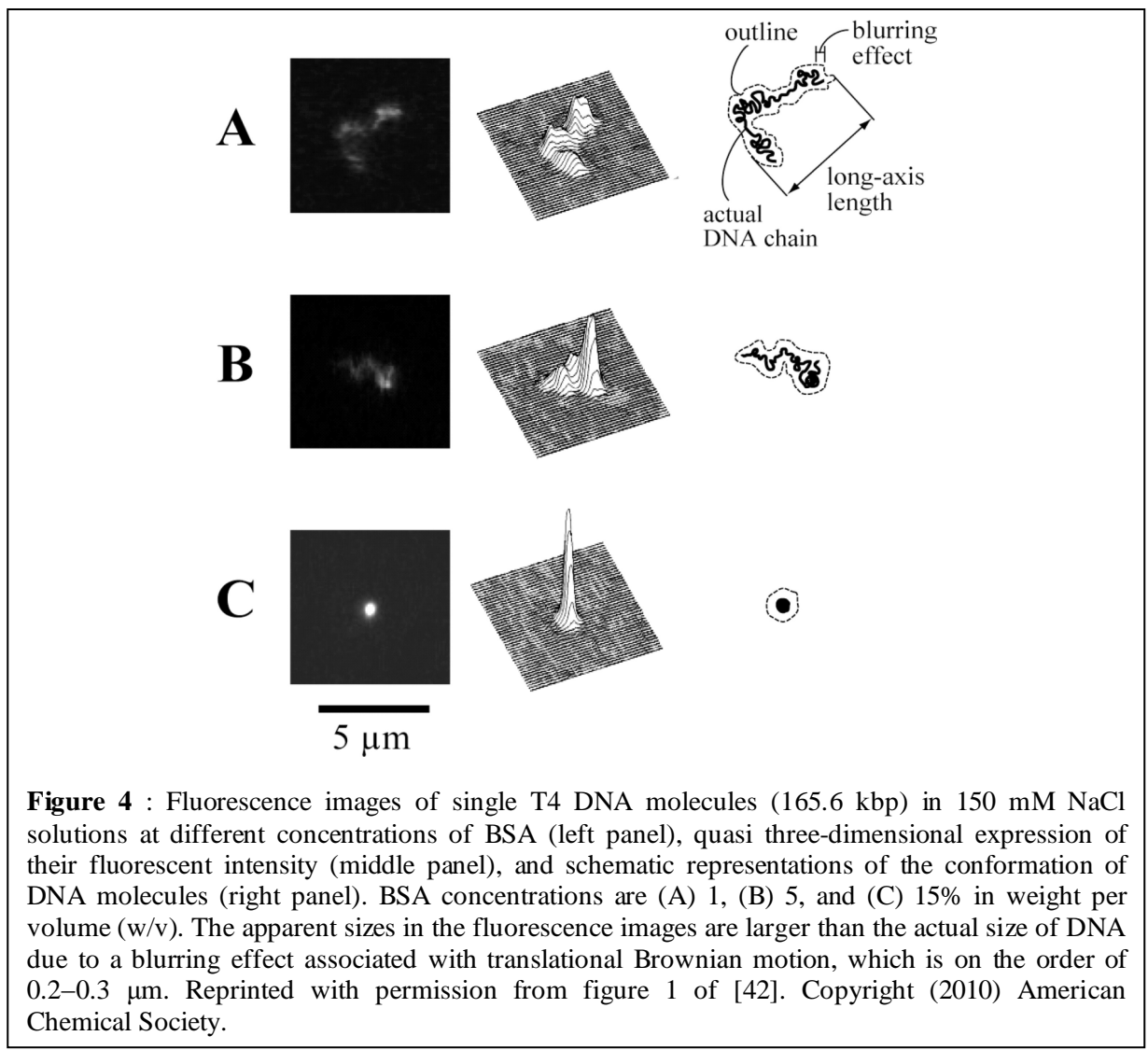

**Figure 4**: Fluorescence images of single T4 DNA molecules (165.6 kbp) in 150 mM NaCl solutions at different concentrations of BSA (left panel), quasi three-dimensional expression of their fluorescent intensity (middle panel), and schematic representations of the conformation of DNA molecules (right panel). BSA concentrations are (A) 1, (B) 5, and (C) 15% in weight per volume (w/v). The apparent sizes in the fluorescence images are larger than the actual size of DNA due to a blurring effect associated with translational Brownian motion, which is on the order of 0.2–0.3 μm. Reprinted with permission from figure 1 of [42]. Copyright (2010) American Chemical Society.

In conclusion, long polycations with concentrations of individual residues equivalent to that of DNA phosphate groups, that is, in the 0.01 M range for physiological conditions, are able to compact DNA efficiently and progressively [38]. No long polycations are, however, known to be present at such large concentrations in the cytoplasm. It is therefore probable that complex coacervation plays no important role in the *in vivo* compaction of bacterial DNA.

*Segregative phase separation.*

In contrast with associative phase separation, which has been the focus of many studies, the possibility that compaction of the nucleoid may result from segregative phase separation has received very little attention up to quite recently. A possible reason for this lack of interest is that it may appear as counter-intuitive that a solution containing macromolecules with like-charges, which consequently all repel each other, may demix and separate into different phases. Still, it is clear that this may happen if, for example, two protein molecules or a protein and a DNA site repel each other more strongly than two DNA sites do. In this case, it is energetically favorable to compact the DNA in a separate region of the cell, so that the average distance between neighboring proteins increases and the strength of the contacts between proteins and between proteins and DNA sites decreases. Segregative phase separation may consequently be expected to occur for solutions containing large concentrations of strongly negatively charged proteins or other macromolecules.

As a matter of fact, segregative phase separation has indeed been recently demonstrated for salt solutions containing approximately 15% (w/v) of the bovine serum albumin (BSA) protein [41,42,43], which is a compact macromolecule ($\approx 4\times4\times14$ nm) with a net negative charge of approximately $-18e$ distributed almost homogeneously on its surface [44]. Results are shown in figure 4. It is seen in this figure that, at 1% BSA and 150 mM NaCl, the 166 kbp bacteriophage T4 DNA exhibits an elongated coil conformation, which is approximately 5 µm long (figure 4(A)), while at 15% BSA the DNA is tightly compacted and appears merely as a bright optical spot in fluorescence microscopy experiments (figure 4(C)). Quite interestingly, at intermediate BSA concentrations, elongated coils and compacted segments coexist along the same DNA molecule (figure 4(B)), thus suggesting that compaction by BSA proteins is a first order transition. Moreover, increasing salt concentration from 150 mM up to 200 mM at 15% BSA results in the DNA decompacting and returning back to a coiled configuration. Addition of salt therefore has a similar effect in solutions containing large amounts of negatively charged BSA proteins as in solutions containing large amounts of long polycations, that is, the increased screening of the charges on the DNA and BSA molecules induced by additional $Na^+$ ions reduces the repulsion between these macromolecules and leads eventually to re-mixing. Note that it was still more recently shown that a few percents of negatively charged silica nanoparticles with diameters ranging from 20 to 135 nm are also able to trigger compaction of DNA from coil to globule configurations [45].

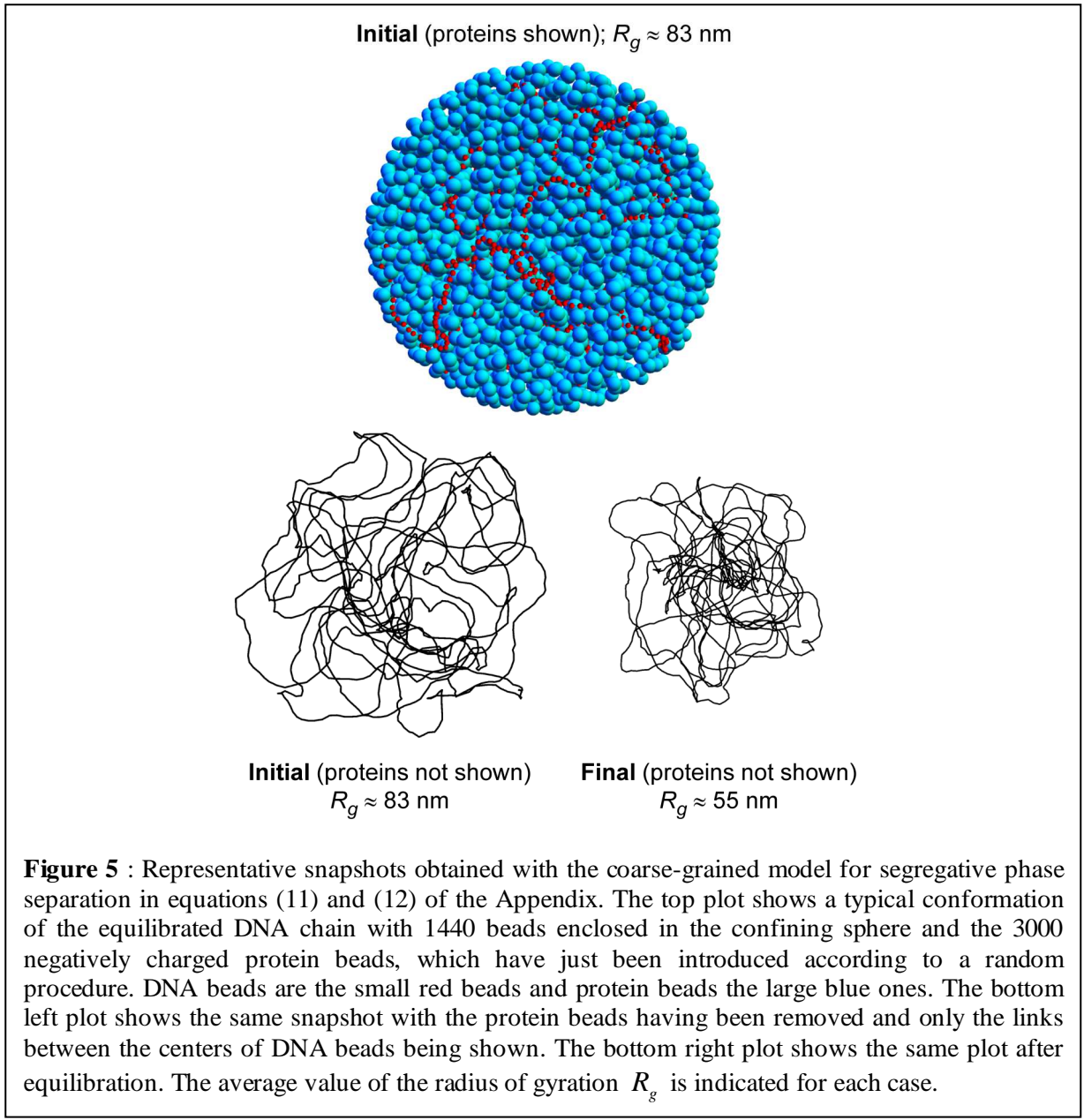

**Figure 5** : Representative snapshots obtained with the coarse-grained model for segregative phase separation in equations (11) and (12) of the Appendix. The top plot shows a typical conformation of the equilibrated DNA chain with 1440 beads enclosed in the confining sphere and the 3000 negatively charged protein beads, which have just been introduced according to a random procedure. DNA beads are the small red beads and protein beads the large blue ones. The bottom left plot shows the same snapshot with the protein beads having been removed and only the links between the centers of DNA beads being shown. The bottom right plot shows the same plot after equilibration. The average value of the radius of gyration $R_g$ is indicated for each case.

Illustrations of segregative phase separation were obtained from simulations performed with a coarse-grained model, which will also be used in the remainder of this Review to discuss the hypotheses that small polycations, supercoiling, and Nucleoid Associated Proteins, may contribute to the compaction of the genomic DNA in the nucleoid. The central part of the model consists of a circular chain of 1440 beads enclosed in a sphere of radius $R_0 = 120$ nm. As in previous work [46,47,48,49,50,51], each bead represents 15 DNA base pairs, has a hydrodynamic radius $a = 1.78$ nm, and is separated at equilibrium from the neighboring ones by a distance $l_0 = 5.0$ nm. The chain therefore represents a sequence with 21600 bp at a concentration close to physiological ones, since both the length of the sequence and the volume of the sphere were obtained by scaling down the values for *E. coli* cells (4.6 Mbp DNA enclosed in a ≈1.5 μm$^3$ cell) by a factor approximately 200. A complete description of the DNA model is provided in the Appendix, equations (1), (2) and (6). After equilibration, the DNA chains adopt coil conformations that fill the whole confining sphere almost homogeneously, as may be seen in the bottom left plot of figure 5. Proteins are introduced in the form of 3000 additional beads with radius $b = 4.8$ nm (protein concentration is 19% v/v), see the top plot of figure 5, which interact with other proteins and with the DNA chain through repulsive electrostatic interactions, as described in equations (11) and (12) of the Appendix. The dynamics of the system composed of the DNA chain and the protein beads is investigated by integrating numerically overdamped Langevin equations, according to equation (15) of the Appendix. The bottom right plot of figure 5 shows that the generalized repulsion between all the beads leads to substantial compaction of DNA, despite the repulsion between neighboring segments and the sterical hindrance of proteins beads. Since quite little is currently known about segregative phase separation induced by generalized electrostatic repulsion, it may be interesting to perform additional simulations with this model to check the respective influences of the size, charge and concentration of the proteins on the degree of DNA compaction.

In conclusion, large amounts of negatively charged macromolecules in the 10-20% v/v concentration range are able to compact DNA efficiently and progressively [41,42]. The concentration of negatively charged macromolecules in prokaryotic cells is difficult to estimate, but it cannot be excluded that it is close to the threshold concentration required for DNA compaction (for example, there are about 100 mg of highly negatively charged RNA per ml of cytoplasm in *E. coli* cells [24]). Segregative phase separation therefore appears as a mechanism that possibly contributes to the formation of the nucleoid in bacteria, although it is only rarely mentioned in the list of possible mechanisms.

**B – Mechanisms based on DNA charge neutralization and/or screening**

In contrast, with associative and segregative phase separation, which involve strong electrostatic interactions between the DNA and other macromolecules present in the cytoplasm to overcome the strong DNA-DNA electrostatic repulsion, the mechanisms discussed in this subsection are instead based on the weakening of the DNA-DNA electrostatic repulsion, which subsequently makes it possible for additional weak forces to provoke DNA compaction. There are essentially two methods for weakening the repulsion between DNA duplexes, namely (i) increase the neutralization of DNA charges by counterion condensation, and (ii) decrease the Debye length.

Manning's theory [52] describes counterion condensation as a phenomenon, which arises from the fact that the ionic atmosphere of counterions with valency $Z$ surrounding a heavily charged polyion immersed in a solvent is unstable when the linear density of effective charges on the polyion is larger than a critical density $\rho_c = e/(Z\ell_B)$, where $\ell_B = e^2/(4\pi\varepsilon_0\varepsilon_r k_B T)$ is the Bjerrum length, that is, the separation at which the electrostatic interaction between two elementary charges $e$ is equal to thermal energy $k_B T$. As a consequence of this instability, counterions coalesce on the polyion and neutralize an increasing number of its charges, till the effective density reduces to the critical value $\rho_c$. The relative dielectric constant of water at 25°C being close to $\varepsilon_r \approx 80$, the Bjerrum length of this medium is of the order of $\ell_B \approx 0.7$ nm. On the other side, as mentioned above, the linear charge density of bare DNA is of the order of $-e$ per 0.17 nm. As a consequence, the neutralization rate is $1 - 0.17/(0.7Z)$, that is approximately 76%, 88%, 92%, and 94% for counterions with valences $Z=1$, 2, 3, and 4, respectively. The cytoplasm contains usually counterions with different valences, which compete for counterion condensation, and more complex formulae have been derived for estimating the total fraction of neutralized DNA phosphate charges when the buffer contains two cations with different valences, see [23,53,54].

On the other side, the Debye (or Debye-Hückel) length $\lambda_D$ is a constant that appears in the linearized Poisson-Boltzmann equation and sets the scale for the variations in the electric potential as well as the concentration of charged species. The shorter the Debye length, the shorter the distance at which a given electrostatic charge is able to exert an influence. The Debye length is equal to $\lambda_D = \sqrt{(\varepsilon_0\varepsilon_r k_B T)/(2N_A e^2 I)}$, where $N_A$ is the Avogadro number and $I = \frac{1}{2}\sum_i c_i z_i^2$ the ionic strength of the solution containing species $i$ with molar concentration $c_i$ (in mole/m$^3$) and charge $z_i$ (in units of the elementary charge $e$). The Debye length is close to 0.97 nm (respectively, 3.1 nm) for an aqueous solution at 25°C containing a monovalent salt at 0.1 M (respectively, 0.01 M) concentration.

It is interesting to consider the effect of adding salt to a solution containing already two different salts, a monovalent one and a salt with valency $Z > 1$. Whatever its valency, addition of salt increases the ionic strength $I$ of the solution and decreases the Debye length. Consequently, screening of the charges carried by the DNA and the other macromolecules becomes more efficient and the effect of the charges less noticeable. Equations in [23,53,54] show that addition of a salt with valency $Z$ (or higher) to the solution similarly increases counterion condensation, because counterions with higher valency replace monovalent ions on the DNA, and consequently decreases the net effective charge on the DNA. In contrast, addition of a monovalent salt decreases counterion condensation, because more monovalent ions compete with multivalent ones for condensation, and consequently increases the net effective charge on the DNA. This remark may help discriminate between different mechanisms, like for example segregative phase separation and macromolecular crowding (see below).

*Condensation by small cations.*

As mentioned above, the efficiency of counterion condensation and DNA charge neutralization increases with the valency of the counterions. In line with this remark, it was shown about 40 years ago that it is possible to condense DNA into very compact structures by adding polyamines with three or four positive charges to very dilute solutions of DNA [55,56,57]. Since that time, this phenomenon has been investigated by many polymer physicists (as a model for the behavior of polyelectrolytes in the presence of multivalent counterions) and biophysicists (as a model for DNA packaging into viruses) and it is now well established that the size and morphology (toroids, rods, or spheres) of condensed DNA depends crucially on the ionic strength and solvent polarity of the solution, as well as the charge and density of the condensing agent (see for example [58,59,60] and references therein). A remarkable result has been obtained by Wilson and Bloomfield, who studied the compaction of T7 bacteriophage DNA by trivalent spermidine, tetravalent spermine and other multivalent cations, at varying monovalent salt concentrations [54]. They were able to shown that, despite the differences between the systems they investigated, condensation uniformly occurred when approximately 90% of the DNA charge was neutralized [54], the amount of neutralization being estimated according to Manning's counterion condensation theory [52,53]. Since the maximum charge neutralization that can be achieved with divalent cations is only 88% (see above), this finding explains why condensation by divalent cations is only marginal.

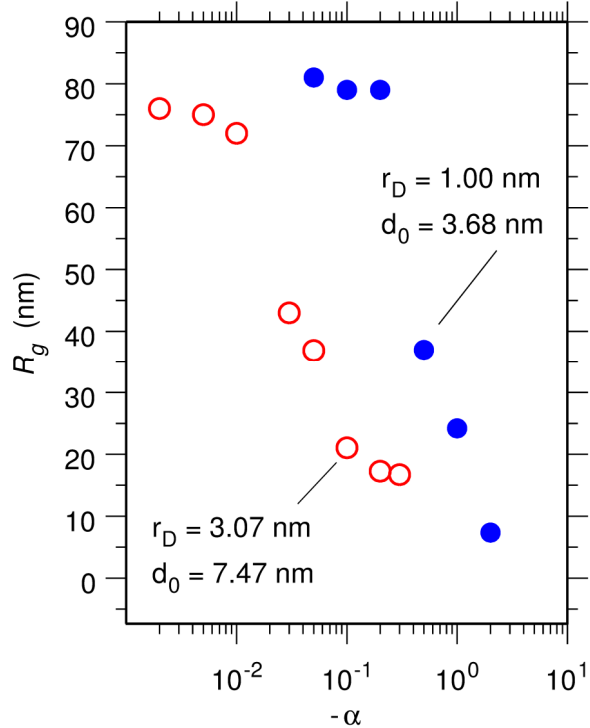

**Figure 6** : Plot of the average radius of gyration $R_g$ of compacted DNA chains with 1440 beads enclosed in a confining sphere as a function of the parameter $-\alpha$ that governs the depth of the DNA-DNA attractive interaction potential (see equation (2) in the Appendix). Empty red circles correspond to results obtained with a broad interaction potential ($r_D = 3.07$ nm and $d_0 \approx 7.47$ in equation (2), see the solid red curve in figure 14) and filled blue dots to results obtained with a narrow attractive potential ($r_D = 1.00$ nm and $d_0 \approx 3.68$ nm in equation (2), see the dashed blue curve in figure 14). Each point was obtained by averaging over four simulations with different initial conditions.

Charge neutralization is therefore fundamental in the condensation of DNA by small cations, because of the associated reduction of the electrostatic repulsion between DNA duplexes, but it cannot by itself induce attraction between the helices and stabilize DNA in the condensed form. A second force is therefore necessarily involved in the condensation of DNA by small cations. Oosawa [61] has shown almost half a century ago that the correlations between the fluctuations of the ionic atmospheres condensed around each rod lead to an attractive force between like-charged rodlike macroions and that this force may eventually be larger than the (partially neutralized) electrostatic repulsion force between the rods if the valency of the counterions is sufficiently large. Marquet and Houssier have later elaborated on this theory to derive an expression for the attractive fluctuation correlation free energy [62]. This expression leads to estimates of the fluctuation correlation energy of the order of -0.3 $k_B T$ per base pair at the condensation threshold, a value which is indeed sufficient to overcome the residual 0.2 $k_B T$ per base pair electrostatic repulsion energy between 90% neutralized DNA helices. DNA condensation by small cations therefore likely results from the conjugate effects of charge neutralization and fluctuation correlation forces.

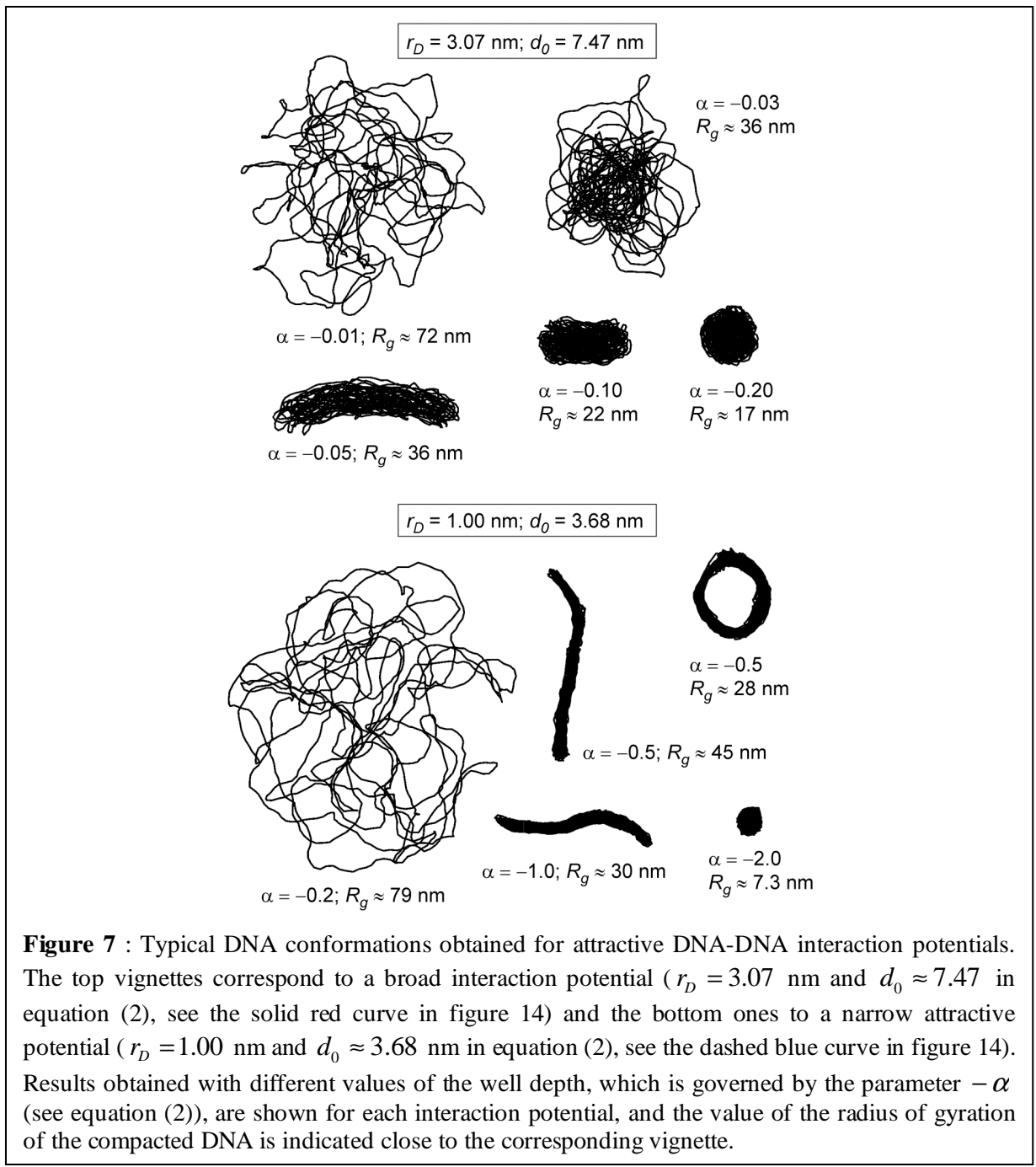

**Figure 7**: Typical DNA conformations obtained for attractive DNA-DNA interaction potentials. The top vignettes correspond to a broad interaction potential ($r_D = 3.07$ nm and $d_0 \approx 7.47$ in equation (2), see the solid red curve in figure 14) and the bottom ones to a narrow attractive potential ($r_D = 1.00$ nm and $d_0 \approx 3.68$ nm in equation (2), see the dashed blue curve in figure 14). Results obtained with different values of the well depth, which is governed by the parameter $-\alpha$ (see equation (2)), are shown for each interaction potential, and the value of the radius of gyration of the compacted DNA is indicated close to the corresponding vignette.

It may be worth emphasizing that complex coacervation (discussed in the previous subsection) and condensation by small cations are rather different mechanisms, although the polycations that trigger them may differ only in their respective number of monomers. The coacervation mechanism is indeed much more efficient than forces arising from fluctuation correlations. For example, it was mentioned above that the compaction of DNA by 92-mer poly-L-lysine molecules is complete in the experimental conditions of [38] when the number of lysine residues is approximately equal to the number of phosphate groups. In contrast, condensation by trimer poly-L-lysine molecules requires approximately 100000 times more lysine residues [38]. Moreover, compaction by long polylysines is gradual (see figure 2(B)), with the volume of DNA decreasing continuously from the coil to the globule value with increasing amino acid concentration, while condensation by small cations is essentially an all-or-none process, with the DNA being either in the coil state or the condensed one but not in-between (see figure 2(C)) [37,38,55,56,57].

There have been several attempts to model DNA condensation by small multivalent cations with coarse-grained models, see for example [63,64,65,66,67]. These studies focused, however, almost exclusively on the formation of toroids and rods when starting from very dilute initial conditions. We performed additional simulations to investigate the compaction dynamics of DNA chains when starting from initial concentrations close to those that prevail in the cell. For this purpose, the model consisting of a circular chain of 1440 beads enclosed in a sphere of radius 120 nm described above was adapted as follows. After allowing the system with repulsive DNA-DNA interactions to equilibrate, the repulsive DNA-DNA interaction term was replaced by an attractive one and the system was allowed to equilibrate again by integrating the Langevin equations. The depth and width of the attractive well were varied in order to check their influence on the degree of compaction. Full detail of the model is provided in the Appendix, see particularly equations (2) and (8), and results are summarized in figures 6 and 7. Figure 6 shows the evolution of the average radius of gyration $R_g$ of compacted DNA chains as a function of the parameter $-\alpha$, which governs the depth of the DNA-DNA attractive interaction potential (see equation (2)), for a broad interaction potential (empty red circles) and a narrow one (filled blue dots). It is seen that compaction of the DNA is indeed a steep transition, which occurs over a narrow interval of values of the depth of the interaction potential, and this interval is all the narrower as the interaction well itself is narrower. Figure 7 additionally displays typical DNA conformations observed at the end of the equilibration procedure. It is seen in the top vignettes that for broad interaction potentials, compaction leads essentially to rods, which shorten and become increasingly spherical with increasing depth of the interaction potential. For narrow interaction potentials (bottom vignettes), several different shapes are instead observed close to the compaction threshold, including thin filaments and toroids. With increasing interaction depth, there only remain filaments, which again shorten, thicken, and become progressively spherical.

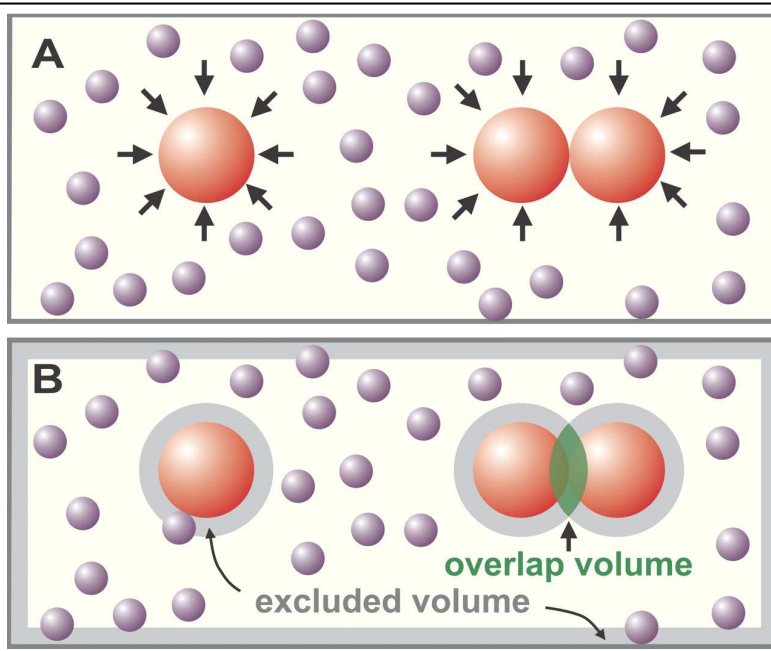

**Figure 8** : Schematic representations of the origin of the depletion force. (A) Many small spheres (purple) bombard three large spheres (red) from all sides (arrows). When two large spheres come into contact (right), the small ones exert a force equivalent to their osmotic pressure on opposite sides of the two large ones to keep them together. (B) The shaded regions in this alternative view show regions inaccessible to the centers of mass of the small spheres. When one large sphere contacts another, their excluded volumes overlap to increase the volume available to the small spheres (increasing their entropy); then aggregation of the large spheres paradoxically increases the entropy of the system. Reprinted from figure 1 of [68].

In conclusion, small polyvalent cations are able to compact DNA abruptly when present at high enough concentrations. Moreover, it is well established that small polyamines like spermidine$^{3+}$ are present in the mM range in bacteria and are efficient in condensing dilute DNA *in vitro*. Still, it is not easy to determine whether such concentrations are sufficient for spermidine to play a role in *in vivo* DNA compaction. Experiments have indeed shown that spermidine concentrations of the order of 100 µM are needed to condense 0.10 µM DNA in 10 mM NaCl solutions [59], that is, a ratio of 1000 between the number of amine residues and phosphate groups. Moreover, increasing the salt concentration up to the 100 mM physiological value certainly tends to make this ratio even larger, since it has been observed that, due to the competition for DNA charge neutralization, the critical concentration of multivalent ions increases with increasing salt [54]. On the other hand, there are some indications that the critical spermidine concentration may increase very sub-linearly with DNA concentration [56]. Current knowledge is consequently insufficient to estimate *a priori* whether millimolar spermidine concentrations are sufficient to play a role in the compaction of 10 mM DNA in 100 mM salt. Still, based on the fact that condensation by small multivalent cations leads to structures with nearly liquid crystalline ordering that are much more compact than the nucleoid of bacteria, where DNA concentration is only of the order of a few percents (w/v), it may be concluded as in [56] that polyamines probably "aid in the packaging of DNA by strongly lowering the free energy of the transition to compact form" but that "they fall short of the conditions for spontaneity".

*Macromolecular crowding, depletion forces, and $\psi$-condensation.*

All mechanisms discussed so far involve charged macromolecules and electrostatic interactions (either attractive or repulsive) between them. In as crowded an environment as the cytoplasm, attractive forces can, however, arise even between hard and chemically and electrically non-interacting species. The mechanism is perhaps most easily understood for mixtures of larger spheres immersed in a liquid containing a very large number of smaller spheres, as illustrated in figure 8. The small spheres move a lot under the influence of thermal forces and bombard the large ones. If the large spheres are far from one another (and far from a wall) the bombardment is isotropic and the resulting force averages to zero. In contrast, if two large spheres are in contact, the smaller spheres cannot approach and hit them from the contact zone. This asymmetry results in a net attractive force between the two large spheres, which tends to keep them in touch. Another more thermodynamical description of this force consists in noticing that there exists around each large sphere a volume from where the center of the smaller spheres is excluded. When two large spheres are in contact, the excluded volumes of the two spheres overlap, so that the total excluded volume decreases and the smaller spheres have access to a larger volume. Stated in other words, aggregation of the large spheres leads somewhat paradoxically to an increase of the entropy of the system. The force arising from the depletion in small spheres of the contact region between the large spheres may therefore be described as an *entropic force*. Asakura and Oosawa were the first to conjecture the existence of this force from theoretical grounds and provide the expression of the depletion force for several case systems [69,70].

Lerman later showed in 1971 that it is indeed possible to induce the collapse of phage DNA molecules into globules approaching the density of phage heads by increasing the concentrations of salt and simple neutral polymers, like poly(ethylene oxide) or poly(ethylene glycol), above certain thresholds, typically 40 mg/ml poly(ethylene glycol) at 1 M NaCl concentration [71]. This discovery, which took place slightly before that of the compaction of DNA by long polyvalent cations (1975, [31]) and small polyvalent cations (1976, [55]) and long before that of the compaction of DNA by negatively charged globular proteins (2010, [41]), triggered an enormous interest from both theoreticians (see for example [72]) and experimentalists and this mechanism is now broadly known as *ψ-condensation*, for Polymer and Salt Induced condensation. The depletion force is indeed rather weak and short-ranged and a high salt concentration is mandatory to reduce the repulsion between charged DNA duplexes. The depletion force does not, however, involve electrostatics and is not screened by salt, in contrast with the three mechanisms described above. It is therefore quite important to realize that addition of monovalent or divalent salt *reduces* the amount of polymer, which is necessary to induce ψ-condensation, while it *increases* the threshold concentrations in long polycations, negatively charged proteins, and small polycations involved in the three other mechanisms. This fact has probably not been recognized clearly enough, since DNA compaction observed in new systems is too often ascribed to ψ-condensation, even when an increase in monovalent or divalent salt concentration has an effect opposite to that expected for ψ-condensation and the compaction mechanism has consequently to be sought elsewhere (see for example [41,42]).

Macromolecular crowding and depletion forces are, however, probably not able to compact the genomic DNA of prokaryotic cells by themselves. This conclusion is suggested by the experiments of Murphy and Zimmerman [6], who showed that upon separation of cell extracts into a DNA-binding fraction and a non-DNA-binding fraction, none of the two fractions is able to induce DNA compaction, while they still do when rejoined. It was later proposed on the basis of theoretical arguments, that the failure of non-DNA-binding extracts to induce DNA compaction may be due to the fact that non-binding macromolecules of the cytoplasm consist mostly of globular proteins and that depletion forces induced by such globular proteins are substantially weaker than those induced by long polymers [73,74], because of the low translational entropy of long polymers. Moreover, ψ-condensation leads to structures with nearly liquid crystalline ordering that are much more compact than the nucleoid of bacteria, like condensation by small polyvalent cations does. Depletion forces are therefore most likely not the primary cause of the compaction of bacterial genomic DNA.

Since the origin of the attractive force between two DNA duplexes was not explicited in the model of DNA condensation by small polyvalent cations discussed in the previous subsection, this model applies equally well to the condensation by depletion forces. These latter forces being, however, particularly short-ranged, relevant simulations should probably be performed with an attraction well substantially narrower than the two ones used in the previous subsection, leading consequently to a transition still steeper than those observed in figure 6.

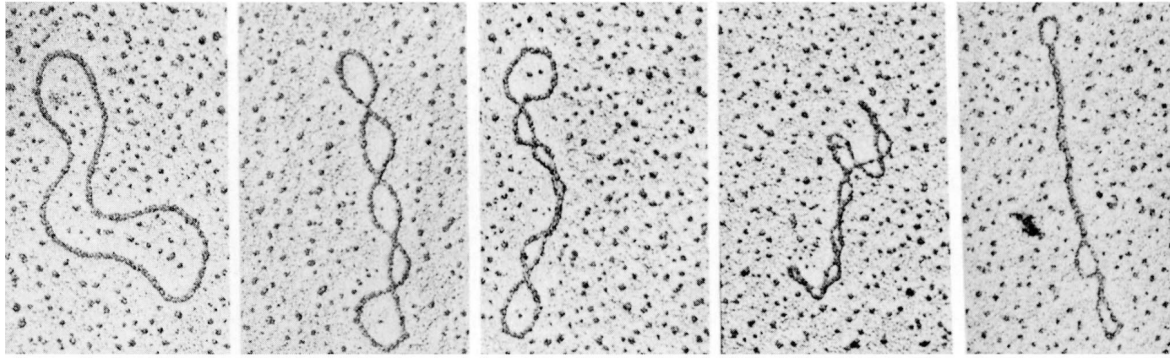

**Figure 9** : Electron micrographs of supercoiled plasmid DNAs. The molecule in the leftmost micrograph is relaxed. The degree of supercoiling, $|\sigma|$, increases from left to right for the three other micrographs. Reproduced from p. 36 of [75], with permission from University Science Books (Mill Valley, California).

**3 – Compaction mechanisms specific to DNA**

Compared to more classical charged polymers, DNA molecules display additional unique properties, among which supercoiling and the ability to bind a large variety of Nucleoid Associated Proteins (NAP) that may in turn influence strongly the geometry of the DNA. In this section, I will discuss the possibility that these two properties contribute to the compaction of the genomic DNA of bacteria. Since NAP represent an important fraction of the proteins found in the cytoplasm and sometimes have a net charge at neutral pH, they may contribute to several of the mechanisms discussed in the previous section, like segregative phase separation and depletion forces. The discussion below deals, however, with possible compaction mechanisms relying on the detail of their specific interactions with DNA, while the mechanisms discussed in the previous section were instead based on non-specific interactions depending uniquely on their volume and global charge.

**A – Supercoiling**

The two strands of a relaxed segment of B-DNA twist around the helical axis and complete a circle once every 10.5 base pairs, a value which depends slightly on the sequence. The number of base pairs of the segment divided by 10.5 is called the linking number and characterizes the number of turns (or twists) along the segment. By looping, cutting, and moving the strands before releasing the cut, some enzymes like topoisomerases are able to add or remove turns from circular (closed) DNA segments, thereby changing the linking number from the value $Lk_0$ of the relaxed segment to a different value $Lk$. The circular DNA is overtwisted if $Lk > Lk_0$ and undertwisted if $Lk < Lk_0$. The quantity $\sigma = (Lk - Lk_0)/Lk_0$ is known as the superhelical density. As a general rule, the DNA of most organisms is undertwisted, with superhelical densities of the order of $\sigma \approx -0.05$ for *E. coli*. Adding or subtracting turns imposes strain in the closed sequence and free circular DNA contorts into more complex shapes to lower this stress. The resulting supercoiled plectonemes are illustrated in figure 9, where $|\sigma|$ increases from left to right. The number of times the double helix crosses over on itself, that is, the number of superhelical twists, is called the writhe $Wr$ ($|Wr| = 4$ in the second left vignette of figure 9). Forming a writhe removes a twist and conversely, so that the sum of the number of twists, $Tw$, and of writhes, $Wr$, remains constant and equal to the linking number $Lk$ during relaxation: $Lk = Wr + Tw$. It has been reported that in 50 mM NaCl solutions, the ratio $Wr/Tw$ remains close to 2.6 over a large interval of values of $\sigma$ [76], which results in one writhe every 290 bp at $\sigma = -0.05$. Moreover, it is believed that the chromosome of *E. coli* may consist of about 400 independently supercoiled domains, each about 10000 base pairs long on average, and with the boundaries between the domains being probably not placed stably along the sequence [4].

Based, maybe, on the observation that individual supercoiled plectonemes formed with short circular DNA molecules like plasmids are more compact than their relaxed counterparts (see for example figure 9), supercoiling is systematically listed as one of the three or four important mechanisms for the compaction of bacterial genomic DNA. There are, however, several indications that the contribution of supercoiling to compaction is probably rather limited. First, it is well known that the enzyme gyrase is essential for the growth of *E. coli* cells and that it consumes ATP to introduce negative supercoiling in the DNA molecule against the increase in potential energy [77]. Still, relaxation of supercoiling by inhibition of the gyrase activity induces only a limited increase in the size of the nucleoid of *E. coli* cells, typically from 5% to 40% in volume [78], while an increase of at least 400% would be needed for the nucleoid to expand through the whole cell. Moreover, it was found that there is only little difference (around 2% w/w) between the critical amounts of poly(ethylene oxide) needed for the ψ-condensation of linear DNA and supercoiled plasmid DNA over a wide range of salt concentrations [79], while a substantially larger difference could reasonably have been expected if supercoiled DNA were significantly compacted compared to linear DNA. Last but not least, the radius of gyration of the genomic DNA of *E. coli* has been estimated by assuming that supercoiled DNA forms a branched polymer [80], adding the contribution of excluded volume interactions [81], and plugging in the value of the branching density of 0.6 branch per kilo-bp obtained by Monte Carlo simulations [82]. These calculations indicate that if supercoiling were the only mechanism for compaction, then the radius of gyration of the nucleoid would be of the order of 3 μm [8], which means that it would not even fit in the cell.

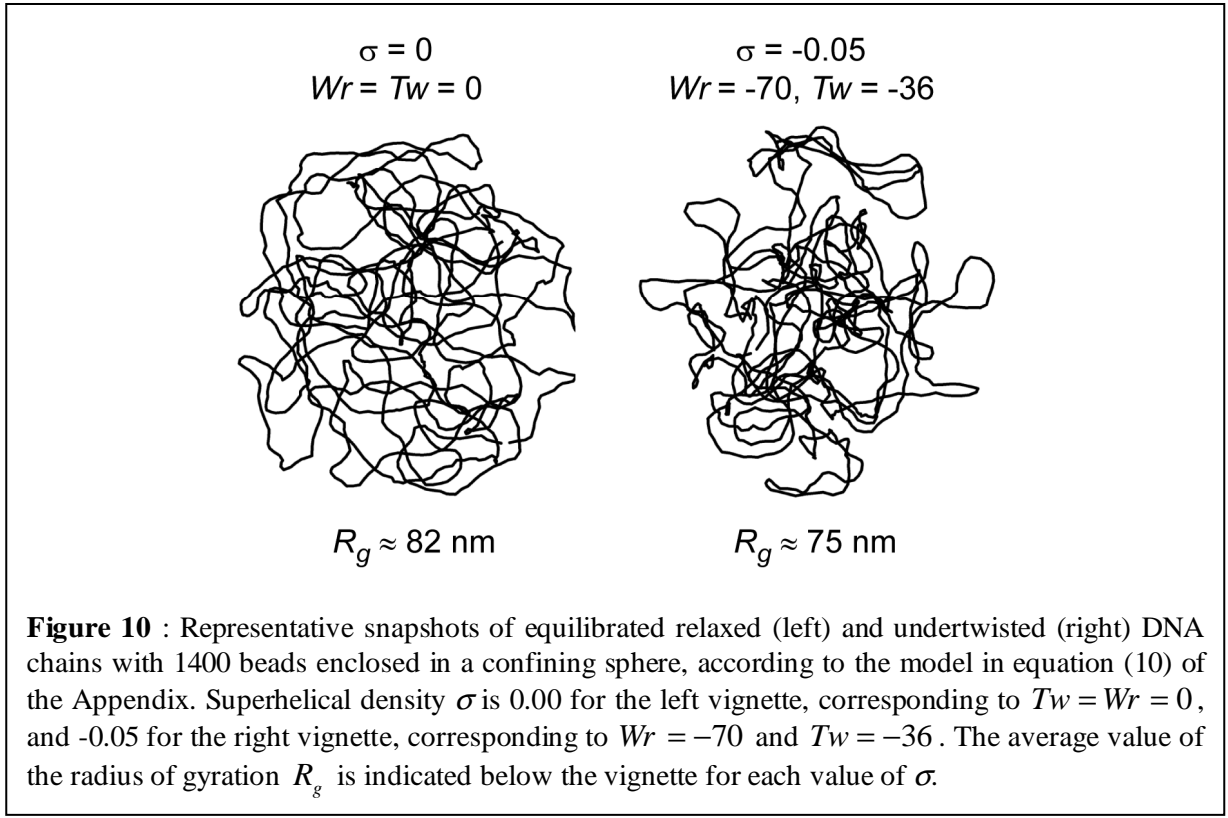

**Figure 10**: Representative snapshots of equilibrated relaxed (left) and undertwisted (right) DNA chains with 1400 beads enclosed in a confining sphere, according to the model in equation (10) of the Appendix. Superhelical density $\sigma$ is 0.00 for the left vignette, corresponding to $Tw = Wr = 0$, and -0.05 for the right vignette, corresponding to $Wr = -70$ and $Tw = -36$. The average value of the radius of gyration $R_g$ is indicated below the vignette for each value of $\sigma$.

The fact that supercoiling leads only to rather mild compaction is confirmed by simulations performed with the coarse-grained model. Torsional degrees of freedom can indeed be introduced in the model according to [83] and spring-spring repulsions according to [84], see equations (9) and (10) of the Appendix. This model predicts that the linking number decomposes approximately into 2/3 of writhe and 1/3 of twist, in fair agreement with experimental measurements [76]. A typical conformation of the DNA chain with 1440 beads enclosed in the confining sphere at superhelical density $\sigma = -0.05$, which is the average value of $\sigma$ for *E. coli* cells, is shown in the right vignette of figure 10. Comparison of this plot with the left plot of the same figure, which shows a typical conformation for relaxed DNA ($\sigma = 0$), illustrates clearly the mild character of the compaction induced by supercoiloing. On average, the radius of gyration $R_g$ of DNA chains with 1440 beads enclosed in the confining sphere decreases only from 82 nm to a modest 75 nm upon decrease of $\sigma$ from 0 to -0.05. Note that the difference in the visual aspects of supercoiled DNA in figures 9 and 10 results mostly from three points, namely (a) long DNA molecules form branched polymers (with a branching density of the order of 0.6 branch per kilo-bp [82]), while shorter plasmids are linear, (b) the spatial extension of writhe loops and the distance between DNA duplexes at crossing points decrease with increasing salt concentrations, and the simulations reported in figure 10 clearly correspond to a significantly smaller salt concentration than the micrograph in figure 9, and (c) deposition of DNA on the charged surface and subsequent drying have a strong impact on the conformation of the DNA, which transforms from a 3D coil to a 2D flat structure [50,51].

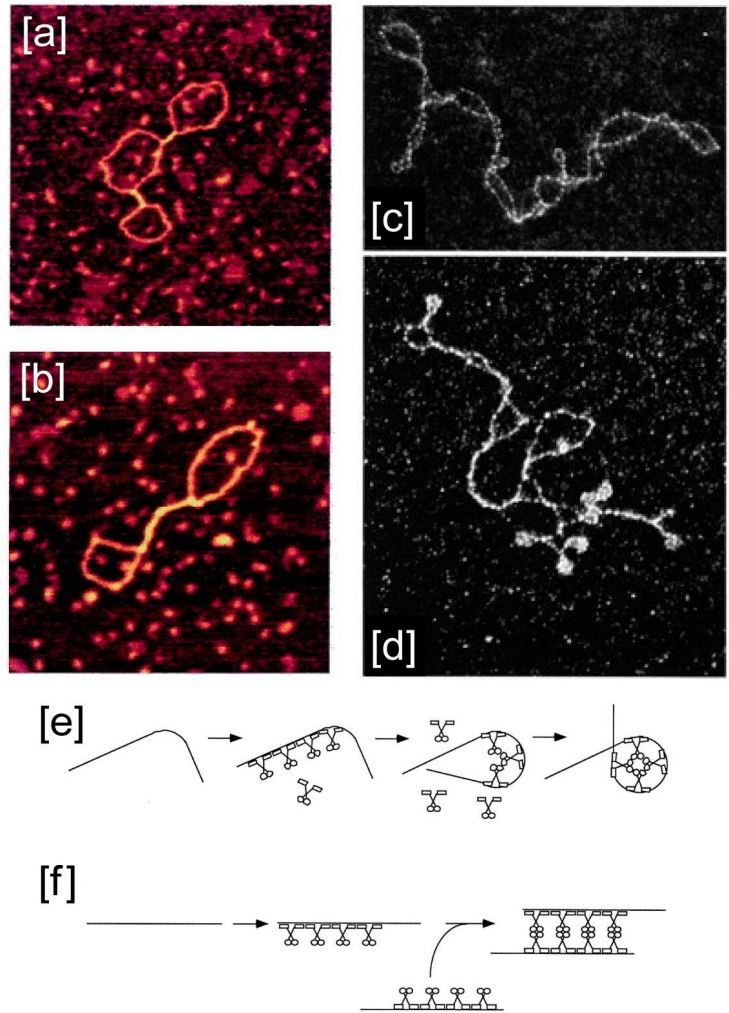

**Figure 11** : (a),(b): AFM images of circular pUC19 after incubation with H-NS proteins, adapted from figure 1 of [9], with permission of Oxford University Press. (c),(d): EM visualization of supercoiled plasmid DNA without protein (c) and after incubation with LrpC proteins (d). (e),(f): Models for LrpC-DNA interaction. (e) Interaction of LrpC with flexible/curved DNA induces a complete wrapping of the DNA around the protein. (f) Interaction of LrpC with straight DNA leads to polymerization of LrpC and bridging of DNA fragments through protein-protein interactions. (c)-(f) are adapted from figures 7 and 8 of [89], with permission from the American Society for Biochemistry and Molecular Biology.

In conclusion, supercoiling certainly contributes to the compaction of bacterial genomic DNA but its contribution is likely to be quite modest and cannot in itself explain the small volume of the nucleoid.

### B – Nucleoid Associated Proteins

As mentioned earlier in this Review, the bacterial nucleoid is essentially composed of nucleic acids and proteins. The major NAP are Fis (factor for inversion stimulation), H-NS (histone-like nucleoid structuring protein), HU (heat-unstable nucleoid protein), IHF (integration host factor protein), Dps (DNA-binding protein from starved cells), and Hfq (host factor for phage $Q_\beta$ replication), but DNA polymerases, RNA polymerase and many species of the transcription factor are also found in the nucleoid [85,86]. The intracellular concentrations of the major protein species vary significantly depending on the growth phase, the most abundant ones in growing cells being Fis, Hfq and HU, while Dps and IHF are predominant during the stationary phase [87]. Some of these proteins have architectural properties, in the sense that they can bridge (like H-NS), bend (like IHF, HU and Fis) or wrap (like Dps) the DNA [88]. DNA bridging by H-NS proteins and its complete wrapping around LrpC (leucine-responsive) proteins are for example clearly seen in figure 11. It is known that the architectural properties of NAP are intimately linked with DNA functions, such as replication, transcription and protection. For example, formation of a bridge between two DNA sites by a H-NS protein may control transcriptional repression by preventing RNA polymerase from binding to a promoter [90] or by trapping it in an inactive open complex [91]. In addition, experiments with mutants strongly suggest that NAP do play a role in the compaction of the nucleoid. For example, bacterial cells lacking both HU and Fis display large decondensed nucleoids [92], while overproduction of H-NS leads in contrast to too dense and compact nucleoids and may be lethal [93]. The hypothesis that NAP may play an important role in DNA compaction is furthermore supported by the fact that at high weight ratios of proteins to DNA, most of the NAP, like Fis [94,95], HU [96,97,98] (referred to as HD in [96]), H-NS [9,94], and IHF [99] induce gradual and strong DNA compaction *in vitro*. Several researchers, however, have pointed out that the intracellular concentrations of NAP are probably substantially smaller than those needed to compact DNA *in vitro*. For HU proteins, it has for instance been estimated that cells contain only from one tenth to one fifth of the number of proteins that is needed to compact the genomic DNA *in vitro* [100,101,102].

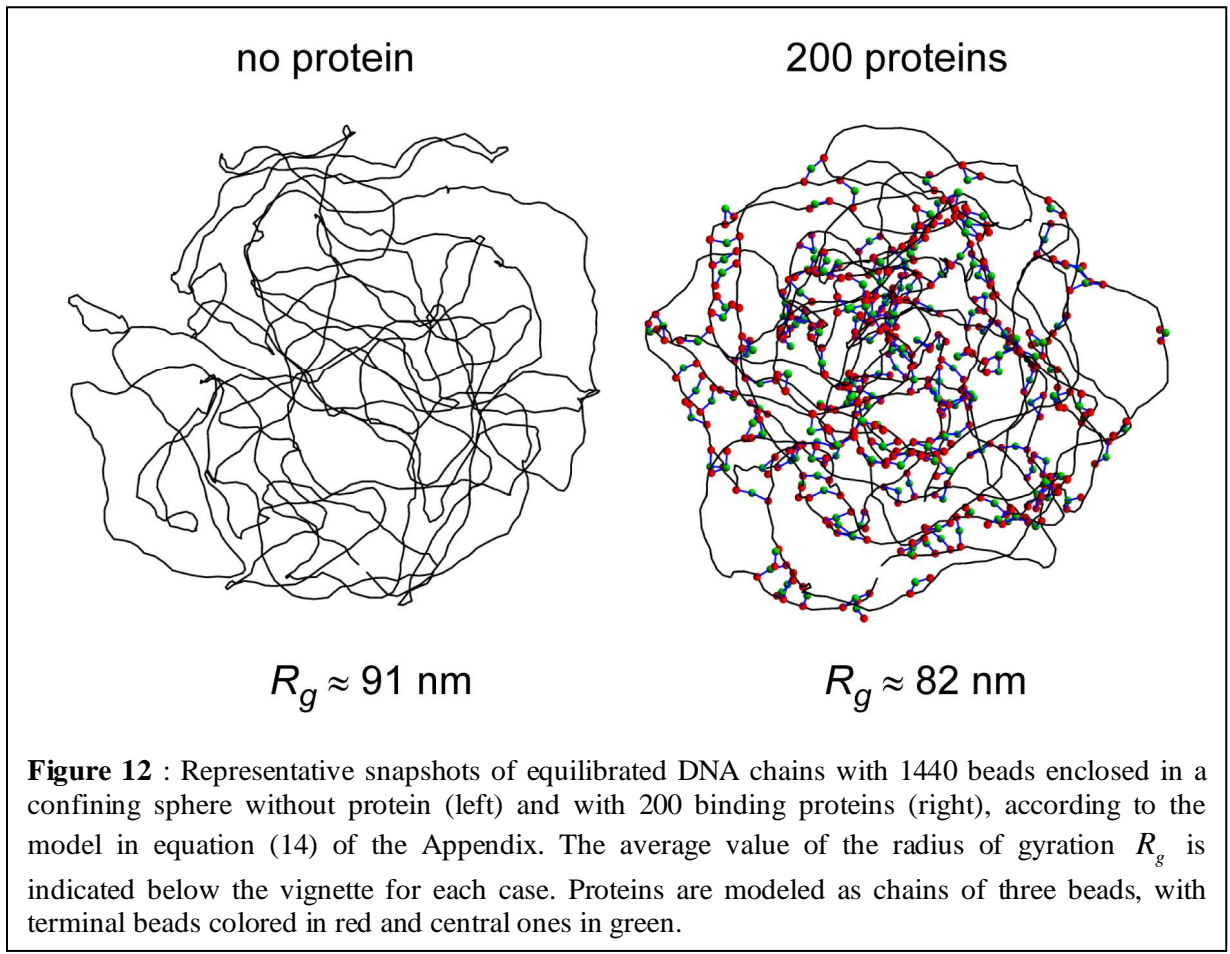

**Figure 12**: Representative snapshots of equilibrated DNA chains with 1440 beads enclosed in a confining sphere without protein (left) and with 200 binding proteins (right), according to the model in equation (14) of the Appendix. The average value of the radius of gyration $R_g$ is indicated below the vignette for each case. Proteins are modeled as chains of three beads, with terminal beads colored in red and central ones in green.

One of the most intriguing questions concerning the eventual role of NAP in the compaction of bacterial genomic DNA is actually whether the putative compaction mechanism is linked to the peculiar architectural properties of the NAP or to other properties that still have been to be determined. For example, it has often been suggested that NAP like H-NS proteins may have a strong influence on DNA compaction because of their ability to form bridges between two DNA duplexes [9,103,104]. If this were the case, then the compaction mechanism would be close to the associative phase separation mechanism discussed in section 2A. The point, however, is that compaction by, say, long poly-L-lysine cations is complete for lysine concentrations approximately equal to the concentration in DNA phosphate groups [37,38]. In contrast, there are at most 26000 H-NS proteins in a *E. coli* cell [87], that is, about one protein for 200 DNA base pairs, which is very likely to be insufficient for compaction. This point can be easily checked using the coarse-grained model for interacting DNA and H-NS molecules, which was proposed in [50,51]. In this model, each H-NS dimer is modeled as a chain of three beads, with the two terminal beads being positively charged and the central one being negatively charged. The interaction between the protein beads and the DNA (modeled as described above) was adjusted so as to reproduce the experimentally determined enthalpy change upon binding of a single H-NS protein to the DNA molecule [105]. Simulations performed with this model were able to reproduce the conformations seen for example in figures 11(a) and 11(b), which consist of plasmids zipped by H-NS bridges, and showed that such zipping most likely originates from the deposition of the DNA-protein complexes on the charged mica surfaces and is probably not important *in vivo* [50,51]. Additional simulations have been performed with this model, starting from equilibrated DNA chains with 1440 beads enclosed in the confining sphere. 200 protein chains were also introduced in the confining sphere, as described in equations (13) and (14) of the Appendix, leading to a protein concentration approximately twice the concentration of H-NS dimers during the cell growth phase and six times the concentration during the stationary phase. Still, equilibrated DNA-protein complexes display only very mild compaction, as can be checked by comparing the two vignettes of figure 12, which show representative snapshots of the DNA with and without proteins. As a matter of fact, introduction of the 200 proteins decreases the average radius of gyration of the DNA chain with 1440 beads from about 91 nm down to a modest 82 nm.

It has also often been suggested that the kinks caused by proteins that are able to bend DNA, like IHF, HU and Fis, may induce significant compaction, but it is actually difficult to imagine why this should be the case. In fact, Atomic Force Microscopy experiments performed with increasing concentrations of HU proteins have shown that the random bends induced by HU proteins at low DNA coverage are responsible for a moderate decrease of the DNA coil size, while large DNA coverage by HU proteins leads instead to an increase of the coil size, because stiff helical DNA-HU fibers are formed [106].

It is therefore likely that the architectural properties of NAP do not by themselves contribute significantly to the compaction of bacterial genomic DNA and that the origin of their compaction ability must be sought elsewhere, for example in their ability to self-associate. In contrast with poly-L-lysine polycations, which are uniformly charged and repel each other like two DNA segments do, most NAP indeed display regions with opposite charges, so that two NAP may usually bind more or less tightly. For example, H-NS is able to bridge two DNA duplexes, because it is functional as a dimer with two terminal DNA-binding domains [9]. Other NAP display a more complex behavior and have different binding modes depending on their concentration and the concentrations in salt and magnesium ions [98,99]. It has for instance been suggested that, in addition to its DNA-bending binding mode, HU may strongly modulate the local DNA geometry by forming nucleosome-like structures, in which a number of HU molecules form an ordered scaffold with DNA lying in the periphery [98]. Similarly, it has been shown that SMC (structural maintenance of chromosomes) proteins are able to form rosette-like structures, in which SMC head domains form the central part of the structures and arms extend outwards, in which DNA loops could be trapped [107]. The influence of such associations of proteins on the compaction of DNA has received little, if any, attention and probably deserves further investigation.

In conclusion, certain NAP certainly do play an important role in the compaction of the nucleoid, but the corresponding mechanism is probably not connected with their architectural properties but rather with other properties yet to be determined, like for example their ability to self-associate.

## 4 – Conclusions and Outlook

In this Topical Review, I have reviewed the mechanisms that may lead to the compaction of the genomic DNA of prokaryotic cells, paying special attention to clarifying the physical interactions they rely upon. These mechanisms may be divided into two broad families, according to whether they are generic to all charged polymers or specific to DNA. Generic mechanisms may again be divided into two sub-families, according to whether they take advantage of the coulombic forces arising from the charges along the DNA or need instead almost complete screening of these charges. These mechanisms are summarized below.

(*) Compaction mechanisms generic to all charged polymers
    (+) Phase separation induced by strong electrostatic interactions
        - Associative phase separation (complex coacervation)
        - Segregative phase separation
    (+) Mechanisms based on DNA charge neutralization and/or screening

- Condensation by small cations

- Macromolecular crowding, depletion forces, and ψ-condensation

(*) Compaction mechanisms specific to DNA

- Supercoiling

- Nucleoid Associated Proteins

As discussed above, the following conclusions hold when considering that each mechanism acts independently from the other ones.
- associative phase separation probably does not play a significant role, because the cytoplasm does not contain sufficient amounts of free long polycations,
- supercoiling is able to induce only limited compaction,
- condensation by small cations and macromolecular crowding are probably not the leading compaction mechanisms, because they would lead to globules that are much more compact than the nucleoid.
- segregative phase separation and Nucleoid Associated Proteins are probably involved in the compaction of the nucleoid but intracellular concentrations of negatively charged macromolecules and NAP are not sufficient to induce compaction *in vitro*.

The general feeling that emerges from this Review is, therefore, that none of the mechanisms listed above is able to compact the bacterial genomic DNA by itself and that (i) either several mechanisms are at play simultaneously, (ii) or the exact mechanism still has to be found.

The hypothesis of the concerted action of at least two mechanisms is supported by several experiments. For example, when extracts of *E. coli* are separated into a DNA-binding fraction and a non-DNA-binding fraction, each fraction taken separately is a poor DNA compacting agent, while they act again similar to the original extracts in the amount required for compaction when rejoined [6]. Moreover, strong DNA compaction is also obtained when neutral polymers are added to the non-DNA-binding fraction [108] or to Nucleoid Associated Proteins [109,110]. On the other hand, it has been conjectured on the basis of a thermodynamic model that the association of negatively charged non-binding proteins and supercoiling is sufficient to induce strong DNA compaction, even in the absence of DNA binding proteins and small multivalent cations [111]. The hypothesis of the concerted action of several mechanisms is further supported by recent experiments, which indicate that there are different levels of nucleoid organization and compaction. More precisely, it has been shown that the nucleoid of *E. coli* cells is composed of four different regions, called macrodomains [13,112,113,114]. Contacts between DNA sites belonging to the same domain are much more frequent than contacts between DNA sites belonging to different domains. While most of the mechanisms described above are able to provide some level of DNA compaction, none of them can account for specific macrodomain organization. It was shown that a 17 kDa protein called MatP is probably responsible for the organization of one of these domains, the so-called Ter macrodomain, which contains the replication terminus [114,115], while it plays no role in the organization and dynamics of the other macrodomains. In the absence of the MatP protein, which binds specifically to the 13 bp *matS* motif repeated 23 times in the 800 kbp Ter macrodomain, DNA in the Ter domain is less compacted, its mobility is larger, and segregation of the Ter domain occurs earlier in the cell cycle. This suggests a multilayer compaction of the nucleoid, with one (or several) of the mechanisms described above inducing general but partial compaction of the DNA, and more specific mechanisms, like the bridging of DNA duplexes by MatP tetramers, being responsible for further compaction and more peculiar organization of the macrodomains.

On the other side, the hypothesis that the most important mechanism for compaction may still have to be found is also supported by several experimental observations. As already mentioned, it has for example been reported that HU proteins may form nucleosome-like structures, in which a number of HU molecules form an ordered scaffold with DNA lying in the periphery [98], and the same conclusion was arrived at for rosettes formed by SMC proteins [107]. The precise influence of such structures on the compaction of bacterial DNA has not been investigated, although it may be expected that the strong electric field generated by such spatially extended complexes may be very efficient in compacting long DNA molecules.

Looking more deeply into possible synergies between the mechanisms listed in this Review and eventually elaborating new ones from experimental observations will therefore probably be the keys for understanding the long standing puzzle of the compaction of the genomic DNA of bacteria into the nucleoid.

**Appendix : Description of the coarse-grained models discussed in this Review.**

The model consists of a circular chain of $n = 1440$ beads with hydrodynamic radius $a = 1.78$ nm separated at equilibrium by a distance $l_0 = 5.0$ nm and enclosed in a sphere with radius $R_0 = 120$ nm, with each bead representing 15 DNA base pairs (see figure 13). This model was obtained by scaling down the length of the genomic DNA and the volume of an *E. coli* cell by a factor of approximately 200, so that the nucleic acid concentration is close to the physiological value. It was furthermore shown in [46] that the value used for the hydrodynamic radius *a* leads to correct diffusion coefficients for the DNA chains. Temperature *T* was assumed to be 298 K throughout the study.

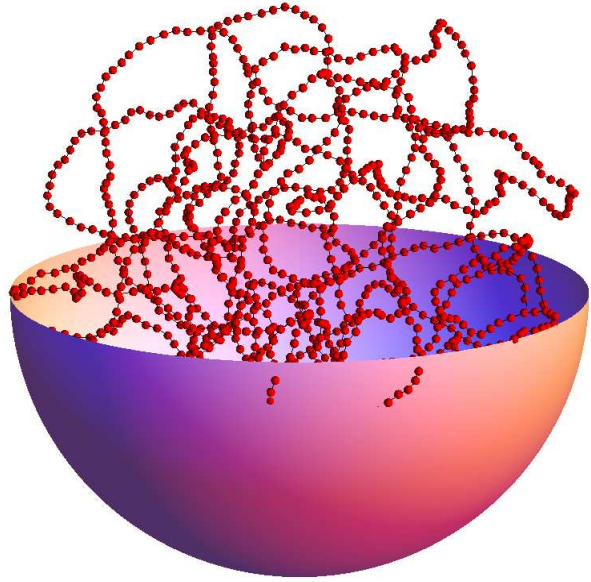

**Figure 13**: Illustration of the coarse-grained model for DNA discussed throughout the Review. Shown in this figure are the DNA chain with 1440 red beads and the confining sphere with radius 120 nm, of which the upper part has been removed. The figure shows a relaxed DNA conformation for the model with torsional degrees of freedom and $\sigma = 0$ described in equation (10) of the Appendix. For the sake of clarity, DNA beads and the confining sphere are usually not displayed in the figures of this Review, only the links between the centers of DNA beads are shown.

The energy of the DNA chain may consist of up to six terms, namely, the stretching energy $V_s$, the bending energy $V_b$, the torsional energy $V_t$, a term $V_e$ that takes both electrostatic and excluded volume interactions into account, a repulsion term $V_X$ that insures that two DNA segments cannot cross, and a confinement term $V_{wall}$ that acts on DNA beads that tend to exit the sphere and repels them back.

The stretching and bending contributions write

$$V_s = \frac{h}{2} \sum_{k=1}^{n} (l_k - l_0)^2$$
$$V_b = \frac{g}{2} \sum_{k=1}^{n} \theta_k^2, \tag{1}$$

where $\mathbf{r}_k$ denotes the position of DNA bead $k$, $l_k = \|\mathbf{r}_k - \mathbf{r}_{k+1}\|$ the distance between two successive beads and $\theta_k = \arccos((\mathbf{r}_k - \mathbf{r}_{k+1})(\mathbf{r}_{k+1} - \mathbf{r}_{k+2})/(\|\mathbf{r}_k - \mathbf{r}_{k+1}\|\|\mathbf{r}_{k+1} - \mathbf{r}_{k+2}\|))$ the angle formed by three successive beads. The stretching energy $V_s$ is a computational device without any biological meaning, which is aimed at avoiding a rigid rod description. The stretching force constant $h$ must be larger than (or equal to) $100\, k_B T / l_0^2$ to insure small enough variations of the distance between successive beads [46]. In contrast, the bending rigidity constant, $g = 9.82\, k_B T$, is chosen so as to provide the correct persistence length for DNA, which is 50 nm, equivalent to 10 beads [46].

Interactions between DNA beads that are not nearest neighbours along the chain are written as the sum of (attractive or repulsive) electrostatic Debye-Hückel terms and (repulsive) excluded volume terms, with the latter ones contributing only if the corresponding electrostatic interactions are attractive

$$V_e = \sum_{k=1}^{n-2} \sum_{K=k+2}^{n} \alpha e_{DNA}^2 H(r_D | \|\mathbf{r}_k - \mathbf{r}_K\|) + \gamma |\alpha| F(d_0 | \|\mathbf{r}_k - \mathbf{r}_K\|), \tag{2}$$

where $H(\rho | r)$ and $F(\rho | r)$ are functions defined according to

$$H(\rho | r) = \frac{1}{4\pi\varepsilon r} \exp\left(-\frac{r}{\rho}\right) \tag{3}$$

and

if $r \leq \rho$: $\quad F(\rho | r) = \frac{\rho^2}{r^2}(\frac{\rho^2}{r^2} - 2) + 1$ (4)

if $r > \rho$: $\quad F(\rho | r) = 0$,

and $\gamma \neq 0$ only if $\alpha < 0$. Electrostatic interactions between nearest-neighbours are not included in equation (2) because it is considered that they are already accounted for in the stretching and bending terms. $\varepsilon = 80\, \varepsilon_0$ denotes the dielectric constant of the buffer. In previous work [46,47,48,49,50,51], the value of the Debye length $r_D$ was set to 3.07 nm in equation (2), which corresponds to a concentration of monovalent salt ions of 0.01 M. Two different values of $r_D$ were instead used in the present work, namely $r_D = 3.07$ nm and $r_D = 1.00$ nm (or $r_D = 1.07$ nm), with the latter values corresponding to a concentration of monovalent salt ions close to the physiological value of 0.1 M. $e_{DNA} = -12.15\,e$ is the electric charge that in previous work [47,48,49,50,51] was placed at the centre of each DNA bead, according to [116]. In equation (2) this charge is allowed to vary and the interaction to switch from repulsive to attractive thanks to the additional parameter $\alpha$. Finally, $d_0$ denotes the threshold distance below which the excluded volume term, described by the repulsive part of a 4$^{th}$ order Lennard-Jones-like function, creates a repulsion force between nearby beads.

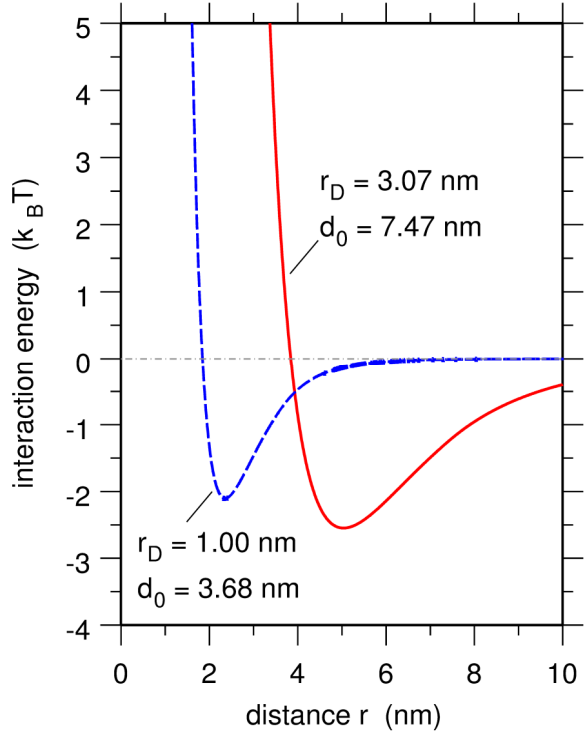

**Figure 14** : Plot, according to equation (2), of the interaction energy of two DNA beads as a function of the separation $r$ between their centers of mass, for $\alpha = -1$, $\gamma = k_B T$, and $r_D = 3.07$ nm and $d_0 \approx 7.47$ nm (solid red curve), or $r_D = 1.00$ nm and $d_0 \approx 3.68$ nm (dashed blue curve).

The additional repulsive term $V_X$ is aimed at preventing two segments from crossing, which sometimes happen for simulations performed with $r_D = 1.00$ nm in equation (2). For this value of the Debye length, the electrostatic repulsion between beads spaced by 5.0 nm along each segment is indeed not always sufficient to prevent two segments from crossing, especially when torsional energy (see below) is large. The segment-segment repulsion term is inspired by the work of Kumar and Larson [84] and writes

$$V_X = \delta \sum_{k=1}^{n-2} \sum_{K=k+2}^{n} F(D_0 | d_{kK}) \;, \tag{5}$$

where $d_{kK}$ is the distance of closest approach of the segments comprised between beads $k$ and $k+1$, on one side, and $K$ and $K+1$, on the other side. It was checked that setting the threshold distance $D_0$ to $1.2\sqrt{2} \approx 1.7$ nm and the prefactor $\delta$ to $k_BT$ or $10 k_BT$ leads to well-behaved repulsion terms, which remain essentially equal to zero for simulations performed with $r_D = 3.07$ nm but nevertheless efficiently prevent segment crossing for simulations performed with $r_D = 1.00$ nm. See [84] for more information on this interaction term and, in particular, how distances of closest approach lying outside the segments are handled.

The confinement term $V_{\text{wall}}$ is taken as a sum of repulsive terms

$$V_{\text{wall}} = \zeta \sum_{j=1}^{n} f(\|\mathbf{r}_j\|) , \tag{6}$$

where $f$ is the function defined according to

if $r \leq R_0$ : $f(r) = 0$

if $r > R_0$ : $f(r) = \left(\dfrac{r}{R_0}\right)^6 - 1$ . $\tag{7}$

The energy function $E_{\text{pot}}$ of the simplest model discussed in this Review, which is introduced in Sect. 2B to illustrate the condensation of DNA by small cations, is the sum of these five terms

$$E_{\text{pot}} = V_s + V_b + V_e + V_X + V_{\text{wall}} , \tag{8}$$

with $h = 100\, k_B T / l_0^2$ in $V_s$, $\gamma = k_B T$ in $V_e$, $\delta = k_B T$ in $V_X$, and $\zeta = k_B T$ in $V_{\text{wall}}$. Two different combinations of the parameters $r_D$ and $d_0$ were furthermore used in equation (2), namely $r_D = 3.07$ nm and $d_0 = 5.28\sqrt{2} \approx 7.47$ nm, and $r_D = 1.00$ nm and $d_0 = 2.60\sqrt{2} \approx 3.68$ nm, in order to check the influence of the value of the bead separation at minimum energy and the width of the attraction well on the shape of the compacted DNA. The curves showing the interaction energy of two beads as a function of the separation $r$ of their center of mass are displayed in figure 14 for $\alpha = -1$ and the two sets of parameters $r_D$ and $d_0$. Simulations were performed with different negative values of $\alpha$ for each set of parameters $r_D$ and $d_0$ to check the effect of the overall interaction strength on the compaction of DNA.

Description of torsional degrees of freedom and related energy terms and forces is borrowed from the work of Chirico and Langowski [83] and the reader is referred to their work for complete information thereon. The torsional energy is written in the form

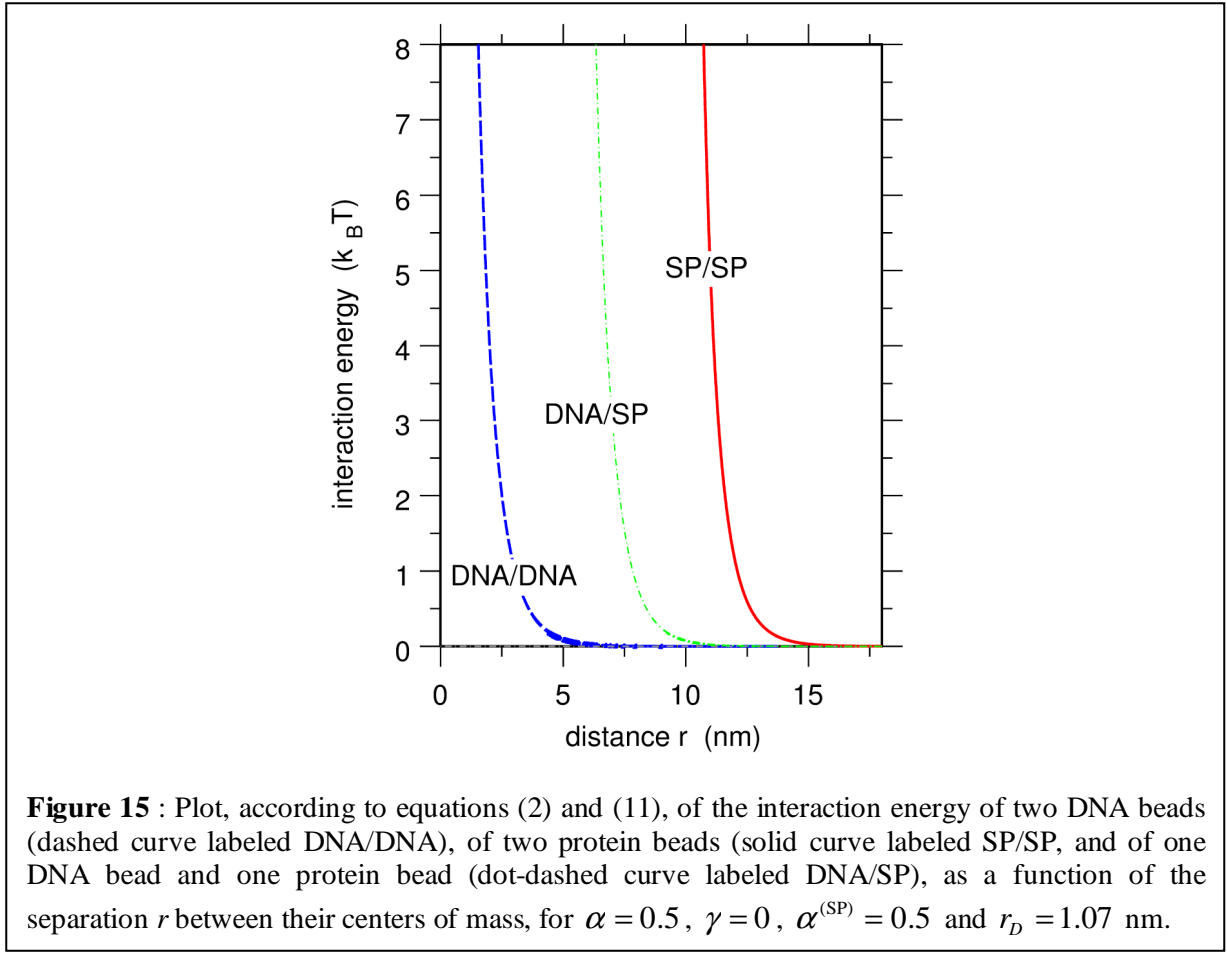

**Figure 15**: Plot, according to equations (2) and (11), of the interaction energy of two DNA beads (dashed curve labeled DNA/DNA), of two protein beads (solid curve labeled SP/SP, and of one DNA bead and one protein bead (dot-dashed curve labeled DNA/SP), as a function of the separation $r$ between their centers of mass, for $\alpha = 0.5$, $\gamma = 0$, $\alpha^{(SP)} = 0.5$ and $r_D = 1.07$ nm.

$$V_t = \frac{\tau}{2} \sum_{k=1}^{n} (\Phi_{k+1} - \Phi_k)^2, \qquad (9)$$

where $\Phi_{k+1} - \Phi_k$ denotes the rotation of the body-fixed frame $(\mathbf{u}_k, \mathbf{f}_k, \mathbf{v}_k)$ between DNA beads $k$ and $k+1$. Published values of the torsion rigidity constant $l_0 \tau$ range from 1.2 10$^{-28}$ to 3.0 10$^{-28}$ Jm [117]. An average value $l_0 \tau = 2.0\ 10^{-28}$ Jm was used in the present work, corresponding to $\tau = 9.72\, k_B T$.

The energy function of the model introduced in Sect. 3A to illustrate the compaction of DNA due to supercoiling is consequently the sum of six terms

$$E_{pot} = V_s + V_b + V_t + V_e + V_X + V_{wall}, \qquad (10)$$

with $h = 100\, k_B T / l_0^2$ in $V_s$, $\alpha = 1$, $\gamma = 0$ and $r_D = 1.07$ nm in $V_e$ (implying that the interaction between DNA beads is repulsive), $\delta = 10\, k_B T$ in $V_X$, and $\zeta = k_B T$ in $V_{wall}$.

Negatively charged spherical proteins are introduced in the model in the form of $N = 3000$ additional beads with radius $b = 4.8$ nm. Proteins therefore occupy a volume fraction of 19% when taking into account only the infinite hard core repulsion (see below).

These beads interact with themselves, with DNA beads, and with the wall of the sphere, so that three additional terms, $V_{\text{e}}^{(\text{SP/SP})}$, $V_{\text{e}}^{(\text{DNA/SP})}$, and $V_{\text{wall}}^{(\text{SP})}$, respectively, must be added to the energy function of the system. They are written in the form

$$V_{\text{e}}^{(\text{SP/SP})} = \sum_{k=1}^{N-1} \sum_{K=k+1}^{N} (\alpha^{(\text{SP})} e_{\text{DNA}})^2 H(r_D \|\mathbf{R}_k - \mathbf{R}_K\| - 2b)$$

$$V_{\text{e}}^{(\text{DNA/SP})} = \sum_{j=1}^{n} \sum_{k=1}^{N} \alpha^{(\text{SP})} e_{\text{DNA}}^2 H(r_D \|\mathbf{r}_j - \mathbf{R}_k\| - b) \qquad (11)$$

$$V_{\text{wall}}^{(\text{SP})} = \zeta^{(\text{SP})} \sum_{k=1}^{N} f(\|\mathbf{R}_k\|),$$

where $\mathbf{R}_k$ denotes the position of protein bead $k$.

The energy function of the model introduced in Sect. 2A to illustrate the compaction of DNA by negatively charged proteins is consequently the sum of seven terms

$$E_{\text{pot}} = V_{\text{s}} + V_{\text{b}} + V_{\text{e}} + V_{\text{wall}} + V_{\text{e}}^{(\text{SP/SP})} + V_{\text{e}}^{(\text{DNA/SP})} + V_{\text{wall}}^{(\text{SP})}, \qquad (12)$$

with $h = 1000\, k_B T / l_0^2$ in $V_{\text{s}}$, $\alpha = 1$ and $\gamma = 0$ in $V_{\text{e}}$ (implying that the interaction between DNA beads is repulsive), $\alpha^{(\text{SP})} = 0.5$ in $V_{\text{e}}^{(\text{SP/SP})}$ and $V_{\text{e}}^{(\text{DNA/SP})}$, $r_D = 1.07$ nm in $V_{\text{e}}$, $V_{\text{e}}^{(\text{SP/SP})}$, and $V_{\text{e}}^{(\text{DNA/SP})}$, $\zeta = 1000\, k_B T$ in $V_{\text{wall}}$, and $\zeta^{(\text{SP})} = 1000\, k_B T$ in $V_{\text{wall}}^{(\text{SP})}$. Note that the strong electrostatic forces exerted by the proteins make it necessary to increase the values of the stretching rigidity $h$ and the strength of the interactions with the wall, $\zeta$ and $\zeta^{(\text{SP})}$, compared to the previous models, in order to keep the distance between successive DNA beads and their distance to the center of the confining sphere at reasonable values. Figure 15 shows the evolution as a function of the distance $r$ between their centers of mass of the interaction energy of two DNA beads (blue dashed line), two protein beads (red solid line), and one DNA bead and one protein bead (dot-dashed green line), according to equations (2) and (11).

Alternatively, Nucleoid Associated Proteins are introduced in the model in the form of chains of three beads with radius $b = a = 1.78$ nm separated at equilibrium by a distance $L_0 = 7.0$ nm, as in [50,51]. Charges $e_{j1} = e_{j3} = -e_{\text{DNA}}/3 \approx 4.05\, e$ are placed at the centre of terminal beads $m = 1$ and $m = 3$ of protein chain $j$ and a charge $e_{j2} = 2e_{\text{DNA}}/3 \approx -8.10\, e$ at the centre of the central bead $m = 2$. The terminal beads of each protein chain are consequently attracted by DNA beads while the central one is repelled, so that each protein chain represents a bridging NAP, like a H-NS dimer or a Lrp octamer. $P = 200$ protein chains are introduced in the confining sphere together with the DNA chain consisting of $n = 1440$ beads, leading to a protein concentration of the model approximately twice the concentration of H-NS dimers during the cell growth phase and six times the concentration during the stationary phase. Protein chains have internal stretching and bending energy and interact with other protein chains, with the DNA chain, and with the sphere wall. Five terms must consequently be added to the energy function of the system. As in [50,51], they are written in the form

$$V_s^{(P)} = \frac{h}{2}\sum_{j=1}^{P}(L_{j,1}-L_0)^2 + (L_{j,2}-L_0)^2$$

$$V_b^{(P)} = \frac{G}{2}\sum_{j=1}^{P}\Theta_j^2$$

$$V_e^{(P/P)} = \sum_{j=1}^{P}\sum_{m=1}^{3}\sum_{J=j+1}^{P}\sum_{M=1}^{3} e_{jm}e_{JM} H(r_D|\|\mathbf{R}_{jm}-\mathbf{R}_{JM}\|)$$
$$+ \chi\sum_{j=1}^{P}\sum_{\substack{J=1\\J\neq j}}^{P} \left(\left|\frac{e_{j1}e_{J2}}{e_{DNA}^2}\right| F(d_0^{(P)}|\|\mathbf{R}_{j1}-\mathbf{R}_{J2}\|) + \left|\frac{e_{j3}e_{J2}}{e_{DNA}^2}\right| F(d_0^{(P)}|\|\mathbf{R}_{j3}-\mathbf{R}_{J2}\|)\right)$$

$$V_e^{(DNA/P)} = \sum_{j=1}^{P}\sum_{m=1}^{3}\sum_{k=1}^{n} e_{jm}e_{DNA} H(r_D|\|\mathbf{R}_{jm}-\mathbf{r}_k\|)$$
$$+ \chi\sum_{j=1}^{P}\sum_{k=1}^{n}\left(\left|\frac{e_{j1}}{e_{DNA}}\right| F(d_0^{(P)}|\|\mathbf{R}_{j1}-\mathbf{r}_k\|) + \left|\frac{e_{j3}}{e_{DNA}}\right| F(d_0^{(P)}|\|\mathbf{R}_{j3}-\mathbf{r}_k\|)\right)$$

$$V_{wall}^{(P)} = \zeta^{(SP)}\sum_{j=1}^{P}\sum_{m=1}^{3} f(\|\mathbf{R}_{jm}\|)$$

(13)

where $\mathbf{R}_{jm}$ denotes the position of bead $m$ of protein chain $j$, $L_{j,1}$ and $L_{j,2}$ the distances between the terminal beads and the central one, and $\Theta_j$ the angle formed by the three beads of protein chain $j$. A value of the bending rigidity $G = 3 k_B T$ was assumed in the simulations. Like $V_e$, the interaction between protein chains, $V_e^{(P/P)}$, and between DNA and protein chains, $V_e^{(DNA/P)}$, is taken as the sum of (attractive or repulsive) electrostatic terms and (repulsive) excluded volume terms, with the latter ones contributing only if the corresponding electrostatic interactions are attractive. Within this model, interactions between DNA and the proteins are essentially driven by the constant $\chi$ in equation (13). The value $\chi = 0.15 k_B T$ was chosen because it leads, for $r_D = 3.07$ nm and $d_0^{(P)} = 2^{3/2}b$, to a change in enthalpy $\Delta H$ of 11.1 $k_B T$ on forming a single bond between DNA and an protein, which is comparable to experimentally determined values for complexes of DNA and H-NS proteins [105]. Attractive interactions between the terminal beads of one protein chain and the central bead of another protein chain are too weak to allow for the formation of long-lived bonds between these two chains, so that formation of stable protein clusters does not occur.

The energy function of the model introduced in Sect. 3B to illustrate the compaction of DNA by bridging NAP is consequently the sum of ten terms

$$E_{\text{pot}} = V_s + V_b + V_e + V_X + V_{\text{wall}} + V_s^{(P)} + V_b^{(P)} + V_e^{(P/P)} + V_e^{(\text{DNA/P})} + V_{\text{wall}}^{(P)} \ , \tag{14}$$

with $h = 100 \, k_B T / l_0^2$ in $V_s$, $\alpha = 1$ and $\gamma = 0$ in $V_e$ (implying that the interaction between DNA beads is repulsive), $r_D = 3.07$ nm in $V_e$, $V_e^{(P/P)}$, and $V_e^{(\text{DNA/P})}$, $\delta = k_B T$ in $V_X$, $\zeta = k_B T$ in $V_{\text{wall}}$, and $\zeta^{(P)} = 10 \, k_B T$ in $V_{\text{wall}}^{(P)}$.

The dynamics of the models described by equations (8), (10), (12) or (14) was investigated by integrating numerically Langevin equations of motion with kinetic energy terms neglected. Practically, the updated position vector for each bead (whether DNA or protein), $\mathbf{r}_j^{(n+1)}$, is computed from the current position vector, $\mathbf{r}_j^{(n)}$, according to

$$\mathbf{r}_j^{(n+1)} = \mathbf{r}_j^{(n)} + \frac{D_t \, \Delta t}{k_B T} \mathbf{F}_j^{(n)} + \sqrt{2 \, D_t \, \Delta t} \, \xi^{(n)} \ , \tag{15}$$

where the translational diffusion coefficient $D_t$ is equal to $(k_B T)/(6\pi\eta a)$ for DNA beads and to $(k_B T)/(6\pi\eta b)$ for protein beads, with $\eta = 0.00089$ Pa s the viscosity of the buffer at $T = 298$ K. $\mathbf{F}_j^{(n)}$ is the vector of inter-particle forces arising from the potential energy $E_{\text{pot}}$, $\xi^{(n)}$ a vector of random numbers extracted at each step $n$ from a Gaussian distribution of mean 0 and variance 1, and $\Delta t$ the integration time step. For the model in equation (10), torsion angles $\Phi_k$ are furthermore updated according to

$$\Phi_k^{(n+1)} = \Phi_k^{(n)} + \frac{D_r \, \Delta t}{k_B T} \tau (\Phi_{k+1}^{(n)} - 2\Phi_k^{(n)} + \Phi_{k-1}^{(n)}) \ , \tag{16}$$

where the rotational diffusion coefficient $D_r$ is equal to $(k_B T)/(4\pi\eta a^2 l_0)$. Time step $\Delta t$ was set to 1 ps for the model in equation (8), to 20 ps for the models in equations (10) and (12), and to 5 ps for the model in equation (14). After each integration step, the position of the centre of the confining sphere was adjusted so as to coincide with the centre of mass of the DNA molecule.